\documentclass[useAMS,usenatbib]{mn2e}
\usepackage{graphicx}
\newcommand{\HI}{H\,{\sc i}}
\newcommand{\HII}{H\,{\sc ii}}
\newcommand{\Ha}{H$\alpha$}
\newcommand{\skms}{\ensuremath{\,\mbox{km}\,\mbox{s}^{-1}}}
\newcommand{\kms}{\ensuremath{\mbox{km}\,\mbox{s}^{-1}}}

\newcommand{\vsys}{\ensuremath{v_{\rm sys}}}

\newcommand{\vrot}{\ensuremath{v_{\rm rot}}}

\newcommand{\LB}{$L_{\rm B}$}

\newcommand{\Msun}{~M$_{\odot}$}

\newcommand{\Lsun}{~L$_{\odot}$}

\newcommand{\AB}{$A_{\rm B}$}

\title[A kinematic study of the gas in IC\,4662 and NGC\,5408]{A kinematic study of the neutral and ionised gas in the irregular dwarf galaxies IC\,4662 and NGC\,5408\thanks{The
    radio observations were obtained with the Australia Telescope which is
    funded by the Commonwealth of Australia for operations as a National
    Facility managed by CSIRO. All optical observations were collected at the
    European Southern Observatory, Chile, Proposal-Nos.: 047.01-003, 51.1-0067,
    69.D-0143(A), and 077B.-0115(A).}}
\author[van Eymeren et al.]{Janine van Eymeren$^{1,2,3}$\thanks{E-mail:
        Janine.Eymeren@rub.de}, B\"arbel
        S. Koribalski$^3$\thanks{E-mail:Baerbel.Koribalski@csiro.au}, \'Angel
        R. L\'opez-S\'anchez$^3$,\newauthor Ralf-J\"urgen Dettmar$^{2}$,
        Dominik J. Bomans$^{2}$
\\
$^1$Jodrell Bank Centre for Astrophysics, School of Physics \& Astronomy, The
University of Manchester, Alan Turing Building,\\Oxford Road, Manchester, M13
9PL, UK\\
$^2$Astronomisches Institut der Ruhr-Universit\"at Bochum,
Universit\"atsstra{\ss}e 150, 44780 Bochum, Germany \\
$^3$Australia Telescope National Facility, CSIRO Astronomy and Space Science,
P.O. Box 76, Epping, NSW 1710, Australia\\
}
\begin{document}

\date{Accepted 2010 April 26. Received 2010 April 22; in original form 2010 February 10}

\pagerange{\pageref{firstpage}--\pageref{lastpage}} \pubyear{2002}

\maketitle

\label{firstpage}

\begin{abstract}
The feedback between massive stars and the interstellar medium is one of the
most important processes in the evolution of dwarf galaxies. This interaction
results in numerous neutral and ionised gas structures that have been found
both in the disc and in the halo of these galaxies. However, their origin and
fate are still poorly understood.\\
We here present new \HI\ and optical data of two Magellanic irregular dwarf
galaxies in the Local Volume: IC\,4662 and NGC\,5408. The \HI\ line data were
obtained with the Australia Telescope Compact Array and are part of the
``Local Volume \HI\ Survey''. They are complemented by optical images and
spectroscopic data obtained with the ESO New Technology Telescope and the ESO
3.6m telescope. Our main aim is to study the kinematics of the neutral and
ionised gas components in order to search for outflowing gas structures and to
make predictions about their fate. Therefore, we perform a Gaussian
decomposition of the \HI\ and \Ha\ line profiles.\\
We find the \HI\ gas envelopes of IC\,4662 and NGC\,5408 to extend well beyond
the optical discs, with \HI\ to optical diameter ratios above four. The
optical disc is embedded into the central \HI\ maximum in both
galaxies. However, higher resolution \HI\ maps show that the \HI\ intensity
peaks are typically offset from the prominent \HII\ regions.\\
While NGC\,5408 shows a fairly regular \HI\ velocity field, which allows us to
derive a rotation curve, IC\,4662 reveals a rather twisted \HI\ velocity
field, possibly caused by a recent merger event. We detect outflows with
velocities between 20 and 60\skms\ in our \Ha\ spectra of both galaxies,
sometimes with \HI\ counterparts of similar velocity. We suggest the existence
of expanding superbubbles, especially in NGC\,5408. This is also
supported by the detection of \emph{FWHMs} as high as 70\skms\ in \Ha, which
cannot be explained by thermal broadening alone. In case of NGC\,5408, we
compare our results with the escape velocity of the galaxy, which shows that
the measured expansion velocities are in all cases too low to allow the gas to
escape from the gravitational potential of NGC\,5408. This result is
consistent with studies of other dwarf galaxies.
\end{abstract}
\section{Introduction}
The interplay between massive stars and the interstellar medium (ISM) has a
large effect on the formation and the evolution of galaxies, especially of
dwarf galaxies.  Photoionisation is the most likely interaction
process. However, stellar winds and supernovae (SNe) explosions of massive
stars also contribute significantly to the energy input into the ISM. The radiative and mechanical feedback heat the gas and drive it outwards, sweeping up the ambient gas into a thin shell. A superbubble filled with hot gas evolves and begins to expand into the ISM. Due
to Rayleigh-Taylor instabilities, the outer shell can rupture and the hot gas can vent out through so-called \emph{chimneys} into the halo
\citep{Norman1989}.

Numerous ionised gas structures such as supergiant shells
or filaments close to the galactic disc of dwarf galaxies, but also at kpc-distances from any place of current star formation, have been detected on
deep \Ha\ images \citep[e.g.,][]{Hunter1993,Bomans1997, Hunter1997,
  Lopez-Sanchez2008a}. Spectroscopic observations of the \Ha\
line revealed that most of the ionised gas structures expand from the disc
into the halo of their host galaxies
\citep[e.g.,][]{Marlowe1995,Martin1998,Bomans2001,vanEymeren2007}. This leads to the question of what the fate of the gas is. One scenario is that the
gas cools down in the halo and eventually falls back onto the galactic disc
\citep[outflow, \emph{galactic fountain} scenario,][]{Shapiro1976}. However,
it might also be possible that the gas escapes from the gravitational
potential by becoming a freely flowing wind (galactic wind). The detection of
large amounts of hot gas in the intergalactic medium (IGM) and the generally
low metal content of dwarf galaxies support this scenario. 

Hydrodynamic simulations modelling superbubble blow-out in dwarf galaxies
\citep{MacLow1999} predict that at least a part of the hot gas has enough
kinetic energy to leave the gravitational potential of its host galaxy and to
enrich the IGM. The relatively low escape velocities of dwarf galaxies should
facilitate the removal of substantial amounts of gas \citep{Larson1974}. According to \citet{Ferrara2000} galactic winds occur in
  dwarf galaxies with gas masses up to $10^9$\,\Msun. However, the fate of the
  gas does not only depend on the mass of the host galaxy and therefore its
  gravitational potential, but it also strongly depends on the morphology of
  the ISM distribution. For a spherical distribution, the ISM seems to be more
  resistant to ejection \citep{Silich2001}.

A recent detailed kinematic study of the neutral and ionised gas in the two
irregular dwarf galaxies NGC\,2366 and NGC\,4861 revealed several outflows in
each galaxy \citep{vanEymeren2009a,vanEymeren2009c}. The measured expansion
velocities were of the order of 20 to 50\skms, which is, in comparison with the
escape velocities of the host galaxies, too low to allow the gas to leave the
gravitational potential. This result confirms earlier observations: no
convincing case of a galactic wind in local dwarf galaxies has been reported
so far \citep{Bomans2005}. Note that galaxies like M\,82 are not typical dwarf
galaxies as they show strong starbursts, are more luminous and have a higher
mass.

Altogether, this shows that despite all the detailed observational
  studies and simulations, we do not fully understand what is going on in these
galaxies. Galactic winds are still thought to be necessary ingredients for
their formation and evolution, but direct evidence seems to be difficult to
  get. In order to improve our knowledge about the processes happening in dwarf
galaxies, we performed a multi-wavelength study of
altogether four irregular dwarf galaxies \citep{vanEymeren2008PhD}. The
results for NGC\,2366 and NGC\,4861 have already been published
\citep[][see also above]{vanEymeren2009a,vanEymeren2009c}. We here concentrate
on the two remaining galaxies IC\,4662 and NGC\,5408. We obtained \HI\ line
data as well as optical images and spectroscopic data. Some basic properties
of both galaxies are given in Table~\ref{basics}.

IC\,4662 (HIPASS J1747--64) is classified as a barred irregular galaxy of
Magellanic type (IBm). Its distance of $D_{\rm TRGB}$\,=\,2.44\,Mpc was
obtained by \citet{Karachentsev2006} and hence makes it the nearest known
representative of blue compact dwarfs \citep{Karachentsev2006}. It seems to be
a rather isolated galaxy, belonging to no known
groups. \citet{deVaucouleurs1975} describes IC\,4662 as a foreground galaxy in
the direction of the NGC\,6300 group.

NGC\,5408 (HIPASS J1403--41) is classified as an IB(s)m galaxy. It was first
studied by \citet{Bohuski1972} who found that its nucleus consists of several
bright \HII\ regions and appears to be undergoing a violent burst of star
formation. As the galaxy reveals an ultra-luminous X-ray source,
NGC\,5408\,X-1, very close to the main \HII\ regions, it was a popular object to study over the last decade
\citep[e.g.,][]{Soria2004,Soria2006,Strohmayer2007,Lang2007,Kaaret2009}.
This X-ray source has recently been argued to harbour an intermediate-mass
black hole \citep{Strohmayer2009}. The distance of NGC\,5408 of $D_{\rm TRGB}$\,=\,4.81\,Mpc was obtained by
\citet{Karachentsev2002}. Its position on the sky puts NGC\,5408 in the
Centaurus A group. The closest known neighbour appears to be ESO\,325-G?001 at
a projected distance of 208\arcmin.\\

This paper is organised as follows: in Sect.~2 we give an overview over the
observations and the data reduction. In Sect.~3 we describe and compare the
morphology of the neutral and ionised gas. Section~4 contains the kinematic
analysis of both gas components including a search for expanding gas. The
results are discussed in Sect.~5, which is followed by a summary in Sect.~6.
\begin{table*}
\centering
\caption{The basic properties of IC\,4662 and NGC\,5408.}
\label{basics} 
$$
\begin{tabular}{lccccc}
\hline
\hline
\noalign{\smallskip}
Optical name         & IC\,4662  & NGC\,5408  & Ref. \\
HIPASS name          & HIPASS\,J1747--64 & HIPASS\,J1403--41 &              \\
\hline
\noalign{\smallskip}
opt. centre:      & & & (1)\\
~~$\alpha$ (J2000)& 17$\rm ^h$ 47$\rm ^m$ 08.8$\rm ^s$ & 14$\rm ^h$ 03$\rm ^m$ 20.9$\rm ^s$ &\\
~~$\delta$ (J2000) & --64\degr\ 38\arcmin\ 30\arcsec & --41\degr\ 22\arcmin\ 40\arcsec & \\
distance [Mpc]       & 2.44          & 4.81        & (2),(3) \\
$v_{\rm opt}$ [\kms]& $336\pm29$    & $537\pm33$  & (1) \\
type                 & IBm           & IB(s)m         & (1) \\
opt. diameter [\arcmin]    & $3.0 \times 1.6$ & $2.6 \times 1.6$ & (4) \\ 
opt. diameter [kpc]             & 2.1 $\times$ 1.1 & 3.6 $\times$ 2.2 \\
inclination [\degr]         &  58 & 52 & (4) \\
position angle [\degr]      & 104 & 62 & (4) \\
\AB\ [mag]           & 0.303 & 0.298 & (5) \\ 
$m_{\rm B}$ [mag]    & $12.33\pm0.09$ & $12.59\pm0.09$ & (4) \\
B-V & 0.41 & 0.56\\
$M_B$ [mag]    & $-14.91\pm0.1$  & $-16.12\pm0.09$ & \\
\LB\ [10$^9$\Lsun]   & $0.14\pm0.01$    & $0.44\pm0.03$ & \\
\noalign{\smallskip}
\hline
\noalign{\smallskip}
$v_{\rm HI}$ [\kms] & $302\pm3$  & $506\pm3$  & (6) \\
$v_{\rm LG}$ [\kms] & 153        & 314        & (6) \\
$w_{\rm 50}$ [\kms] &  86        &  62        & (6) \\
$w_{\rm 20}$ [\kms] & 133        & 112        & (6) \\
$F_{\rm HI}$ [Jy\skms]       & $130.0\pm12.0$ & $61.5\pm6.7$ & (6) \\
$M_{\rm HI}$ [10$^8$\Msun]    & $1.83\pm0.17$  & $3.36\pm0.37$ \\ 
\noalign{\smallskip}
\hline
\hline
\end{tabular}
$$
\flushleft\footnotesize{
Note: The blue luminosity is calculated using a solar $B$-band magnitude of 5.48 mag. -- References: 
  (1) \citet{deVaucouleurs1991}, 
  (2) \citet{Karachentsev2006},
  (3) \citet{Karachentsev2002},
  (4) \citet{Lauberts1989} [ESO Uppsala], 
  (5) \citet{Schlegel1998},
  (6) \citet{Koribalski2004} [HIPASS BGC].}
\end{table*}
\section{Observations and Data Reduction}
\subsection{Optical data}
\subsubsection{Imaging}
\begin{figure*}
\centering
\includegraphics[width=\textwidth]{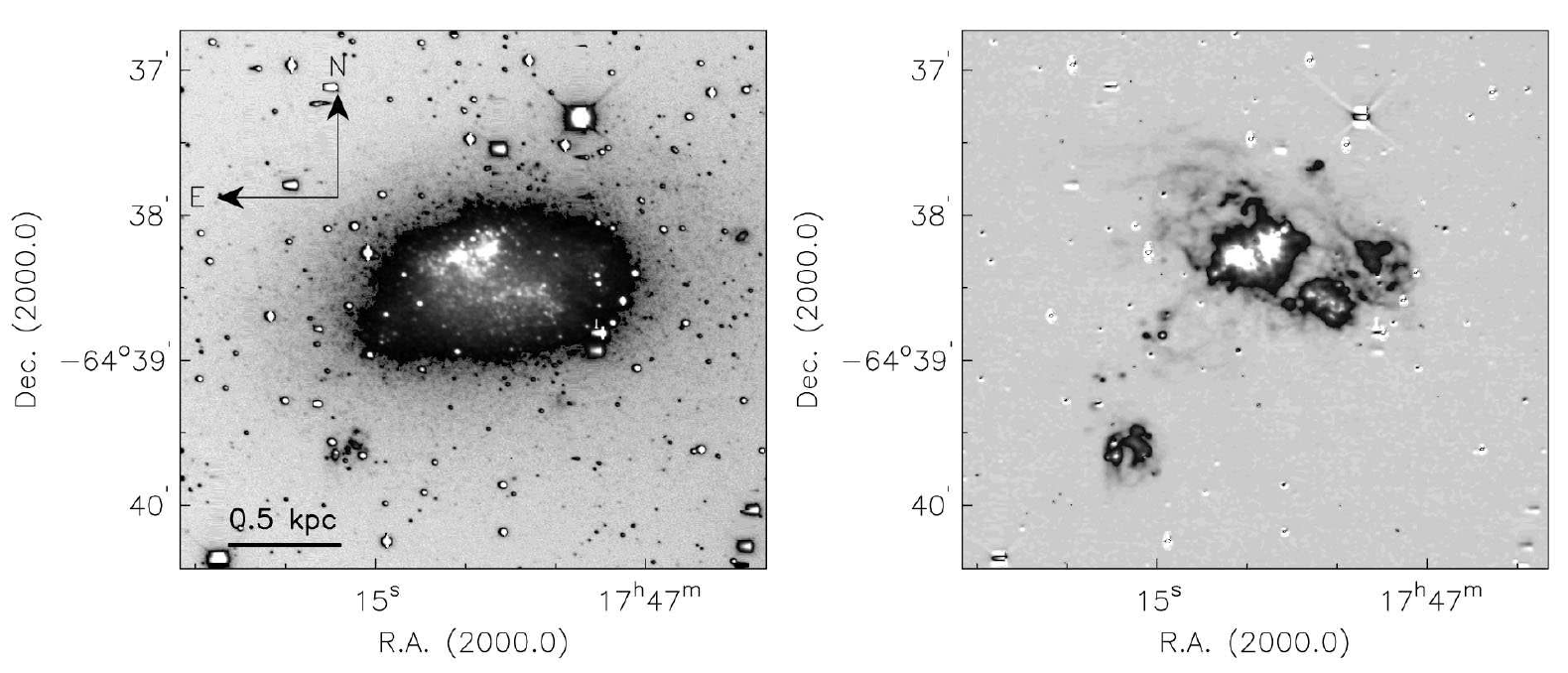}
\includegraphics[width=\textwidth]{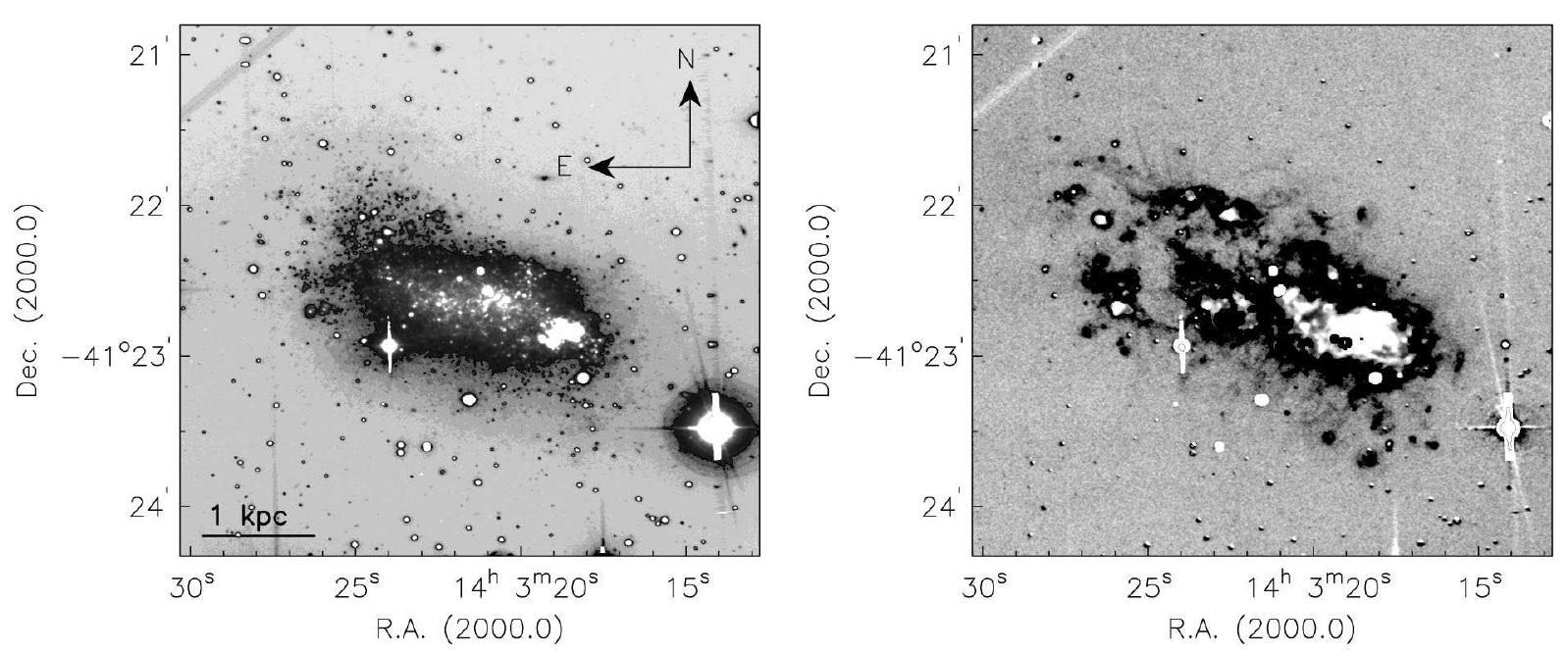}
\caption[\emph{R}-band and \Ha\ images of IC\,4662 (top) and NGC\,5408
  (bottom).]{Optical imaging of the Magellanic irregular dwarf galaxies
  IC\,4662 (top) and NGC\,5408 (bottom). {\bf Left panel:} \emph{R}-band image;
  {\bf right panel:} continuum-subtracted \Ha\ image. The images of IC\,4662
  were obtained with the ESO NTT, the images of NGC\,5408 with the ESO 3.6m
  telescope. In order to locate the positions of the main star
  clusters and the main \HII\ regions respectively, the high intensity areas
  in both images are highlighted in white.}
\label{Rband}
\end{figure*}

Deep \emph{R}-band and \Ha\ images are available for both galaxies. Using the
ESO Multi-Mode Instrument (EMMI) attached to the ESO New Technology Telescope
(NTT), we obtained a 600\,s exposure of IC\,4662 in \emph{R}-band and a
1800\,s exposure in \Ha. A 600\,s \emph{R}-band image of NGC\,5408 as well as
a 1200\,s \Ha\ image, both observed with the ESO Faint Object Spectrograph and
Camera (EFOSC) attached to the ESO 3.6m telescope, have been taken from the
ESO archive. The data reduction was performed using the software package IRAF
\citep{Tody1993}, and included standard procedures of overscan- and
bias-subtraction as well as a flatfield correction. Additionally, we
removed cosmic rays by running the IRAF version of L.~A. Cosmic
\citep{vanDokkum2001}. In order to get the pure \Ha\ line emission, we first
scaled the flux of the stars in both the continuum and the \Ha\ images and
subsequently subtracted the continuum image from the \Ha\ image. The seeing
was 1\farcs1 and 0\farcs8 during the observations of IC\,4662 and NGC\,5408
respectively. We used adaptive filters based on the H-transform algorithm
\citep{Richter1991} to stress weaker structures in the \Ha\ images and to
differentiate them from the noise. All images are displayed in
Fig.~\ref{Rband}. Some observational details are given in
Table~\ref{imageobs}.
\begin{table}
\centering
\caption{Imaging -- some observational parameters.}
\label{imageobs}
$$
\begin{tabular}{lcc}
\hline
\hline
Parameter [Unit] & IC\,4662 & NGC\,5408\\
\hline
Telescope & ESO NTT & ESO 3.6m\\
Instrument & EMMI & EFOSC\\
Filter & 608 (R), 596 (\Ha) & 642 (R), 692 (\Ha) \\
Date & 07.05.91 & 10.07.02\\
Exp. Time [s] & 1$\times$600, 1$\times$1800 & 1$\times$600, 1$\times$1200\\
Spatial res. [\arcsec] & 0.36 & 0.16\\
Airmass & 1.23 & 1.02\\
Seeing [\arcsec] & 1.1 & 0.8\\
\hline
\hline
\end{tabular}
$$
\end{table}
\subsubsection{Medium-resolution long-slit spectroscopy}
\label{medspec}
We used archival medium-resolution long-slit spectra of NGC\,5408, which have a
spectral resolution of about 60\skms, as measured from the night sky lines. As
the target of these observations was the ultra-luminous X-ray source
NGC\,5408\,X-1, all spectra were obtained at almost the same
position. Therefore, we only analyse one slit position as a
representative. The seeing was 0\farcs9.

We performed further medium-resolution long-slit spectroscopy not only of
NGC\,5408, but also of IC\,4662 from May 1st to May 3rd 2006. As we aimed for
very deep spectra and were mainly interested in the \Ha\ line, only the red
part of EMMI was read out. We used grating \#\,7 centred on the \Ha\ line
with a wavelength coverage of 1300\,\AA\ and a dispersion of
0.41\,\AA/pix. With a slit width of 1\arcsec\ we achieved a spectral
resolution of 112\skms, again as measured from the night sky lines. The
pixel size is 0\farcs332 and the seeing was about 1\arcsec. Spectra of a
thorium comparison lamp were taken for the wavelength calibration.

We obtained three spectra each at different positions across the two
galaxies. The slit positions were chosen to intersect prominent shell and
filamentary structures as visible on the \Ha\ images. Some details of the
observations are given in Table~\ref{spectra_exptime}. 

The data reduction of all spectra (including the archival one) was performed
by us using IRAF and included overscan- and bias-subtraction, flatfield
correction and a wavelength calibration. A geometric distortion correction
was not necessary because the deviations were smaller than one pixel. Again,
cosmic rays were removed by running L.~A. Cosmic. We also performed a
background subtraction to remove the contamination by the night sky lines.

The spectra and their analysis are presented in Sect.~\ref{comp_spec}. They
were binned by three pixels in spatial direction, which corresponds to about
1\arcsec\ in order to match the seeing (see above). At positions of very weak
emission, ten pixels were summed up to improve the signal to noise ratio (S/N).
\begin{table*}
\centering
\caption{Long-slit spectroscopy -- some observational details.}
\label{spectra_exptime}
$$
\begin{tabular}{lccccc}
 \hline
  \hline
  Object & Telescope / Instrument & Date & Slit No. & Exp. Time & $PA$\\
 & & & & [min] & [$\degr$]\\
 \hline
 NGC 5408 & ESO NTT / EMMI & 01.05.06 & 1a & 5$\times$45 & 54\\
 & & 02.05.06 & 2 & 4$\times$45 & 6\\
 & & 03.05.06 & 3 & 4$\times$45, 1$\times$21.5 & 296\\
 & & 17.05.93 & 1b & 3$\times$20 & 80\\
 \hline
 IC 4662 & ESO NTT / EMMI & 02.05.06 & 1 & 4$\times$45 & 330\\
 & & 03.05.06 & 2 & 3$\times$45 & 56\\
 & & 02.05.06 & 3 & 4$\times$45 & 70\\
  \hline
  \hline
\end{tabular}
$$
\flushleft\footnotesize{Note: Slit\,1b of NGC\,5408 denotes the archival
  spectrum. The position angle \emph{PA} is defined to increase
  counter-clockwise with north being 0\degr.}
\end{table*}
\subsection{ATCA \HI\ synthesis observations}
\label{sectradiored}
\HI\ line observations of IC\,4662 and NGC\,5408
were obtained with the Australia Telescope Compact Array (ATCA), and are part
of the ``Local Volume \HI\ Survey'' \citep[LVHIS\footnote{LVHIS is a large
    project that aims to provide detailed \HI\ intensity maps, velocity fields
    and 20\,cm radio continuum observations for a complete sample of nearby,
    gas-rich galaxies belonging to the Local Volume (LV), a sphere of radius
    10\,Mpc centred on the Local Group. The ATCA observations include all LV
    galaxies which were detected in HIPASS and reside south of
    $\delta \approx$\,-30\degr. Further details can be found at \emph{http://www.atnf.csiro.au/research/LVHIS/}.};][]{Koribalski2008,Koribalski2010}.

The first frequency band was centred on 1418 MHz with a bandwidth of 8 MHz,
divided into 512 channels. This gives a channel width of 3.3\skms\ and a
velocity resolution of 4\skms. The ATCA primary beam is 33\farcm6 at 1418 MHz,
which is sufficient to fully map both galaxies and their surroundings.

In order to ensure excellent \emph{uv}-coverage and sensitivity to large-scale
structures, IC\,4662 was observed for a full synthesis (12\,h) in each the
EW\,367, 750A, and 1.5D configurations. The lower resolution data of NGC\,5408
(375 and 750D array) were taken from the archive and complemented by
high-resolution data within LVHIS (1.5A array).

The data reduction was carried out with the MIRIAD software package
\citep{Sault1995} using standard procedures, including flux- and phase
calibrations. Using a first order fit to the line-free channels in the
\HI\ data set, the 20\,cm radio continuum was separated from the \HI\
emission. We created two different sets of data cubes, a low-resolution cube to
detect the extended emission and a high-resolution cube to resolve the
inner structure and to make the \HI\ data comparable to the optical data. For
the low-resolution cubes of both galaxies, we excluded the longest baselines,
which are all baselines to the distant antenna~6 (CA06). The low-resolution
data of IC\,4662 were made using 'natural' weighting of the {\em
  uv}-data in the velocity range covered by the \HI\ emission in steps of
4\skms. The low-resolution data of NGC\,5408 were made using 'robust'
weighting. The high-resolution data of both galaxies were created by including
all baselines to CA06 and using the same weighting as for the low-resolution
cubes. The resulting beam sizes are given in Table~\ref{atcaobs}.

1\,Jy\,beam$^{-1}$ corresponds to an \HI\ column density of $2.55\times
10^{20}$ atoms~cm$^{-2}$ (IC\,4662, low-res.), $2.64\times 10^{21}$
atoms~cm$^{-2}$ (IC\,4662, high-res.), $2.89\times 10^{20}$ atoms~cm$^{-2}$
(NGC\,5408, low-res.), and $2.77\times 10^{21}$ atoms~cm$^{-2}$ (NGC\,5408,
high-res.). The moment maps (integrated intensity map, intensity-weighted mean
velocity field, and the velocity dispersion) were created from the \HI\ data
cubes by first isolating the regions of significant emission in every channel
and afterwards clipping everything below a 2.5$\sigma$ threshold. This final
step of the data reduction process and the subsequent analysis of the \HI\
data were performed with The Groningen Image Processing System
\citep[GIPSY,][]{vanderHulst1992}, complemented by some IRAF tasks.
\begin{table}
\centering
\caption{ATCA \HI\ data -- imaging properties.}
\label{atcaobs}
$$
\begin{tabular}{lcc}
\hline
\hline
Object         & IC\,4662  & NGC\,5408   \\
\hline
Low-resolution: & &\\
Weighting & natural, $-$CA06 & robust, $-$CA06\\
\HI\ synthesised beam & 70\arcsec\ $\times$\,62\arcsec &
71\arcsec$\times$\,54\arcsec\\
\hline
High-resolution: & &\\
Weighting & natural, $+$CA06 & robust, $+$CA06\\
\HI\ synthesised beam & 21\arcsec\ $\times$\,20\arcsec & 20\arcsec$\times$\,20\arcsec\\
\hline
\hline
\end{tabular}
$$
\end{table}
\section{General morphology}
\label{LVHISgenmorph}
In order to get a first impression of how the stars and the gas are
distributed, we compared the optical and \HI\ morphologies of IC\,4662 and
NGC\,5408.
\subsection{IC\,4662}
\label{SectI4662morpho}
The \emph{R}-band image of IC\,4662 (see Fig.~\ref{Rband}, upper left panel)
reveals a slightly elongated, box-like shape with a position angle of
104\degr\ (Table~\ref{basics}). The \Ha\ image (Fig.~\ref{Rband}, upper right
panel) shows a large \HII\ region complex with an \Ha\ minimum in the
south-eastern part of the stellar light distribution, and extended diffuse,
filamentary gas structures outside the stellar disc to the north-east. The
central two star clusters are offset by about 350\,pc to the north-east from
the optical centre (see Table~\ref{basics}), which probably explains the
extended \Ha\ emission in this area. They coincide with the two brightest
\HII\ regions visible on the \Ha\ image. Additionally, a star cluster with
associated \Ha\ emission that appears to be detached from the main body of the
galaxy can be seen. It is located 1\farcm5 or 1.1\,kpc to the south-east of
the centre.  Our deep \Ha\ image shows an \HII\ region at the same position
that is connected to the main complex by small, compact \HII\ regions and
diffuse filamentary gas structures. We will come back to this feature in
Sect.~\ref{southHII}.

\begin{figure*}
\centering
\includegraphics[width=.92\textwidth]{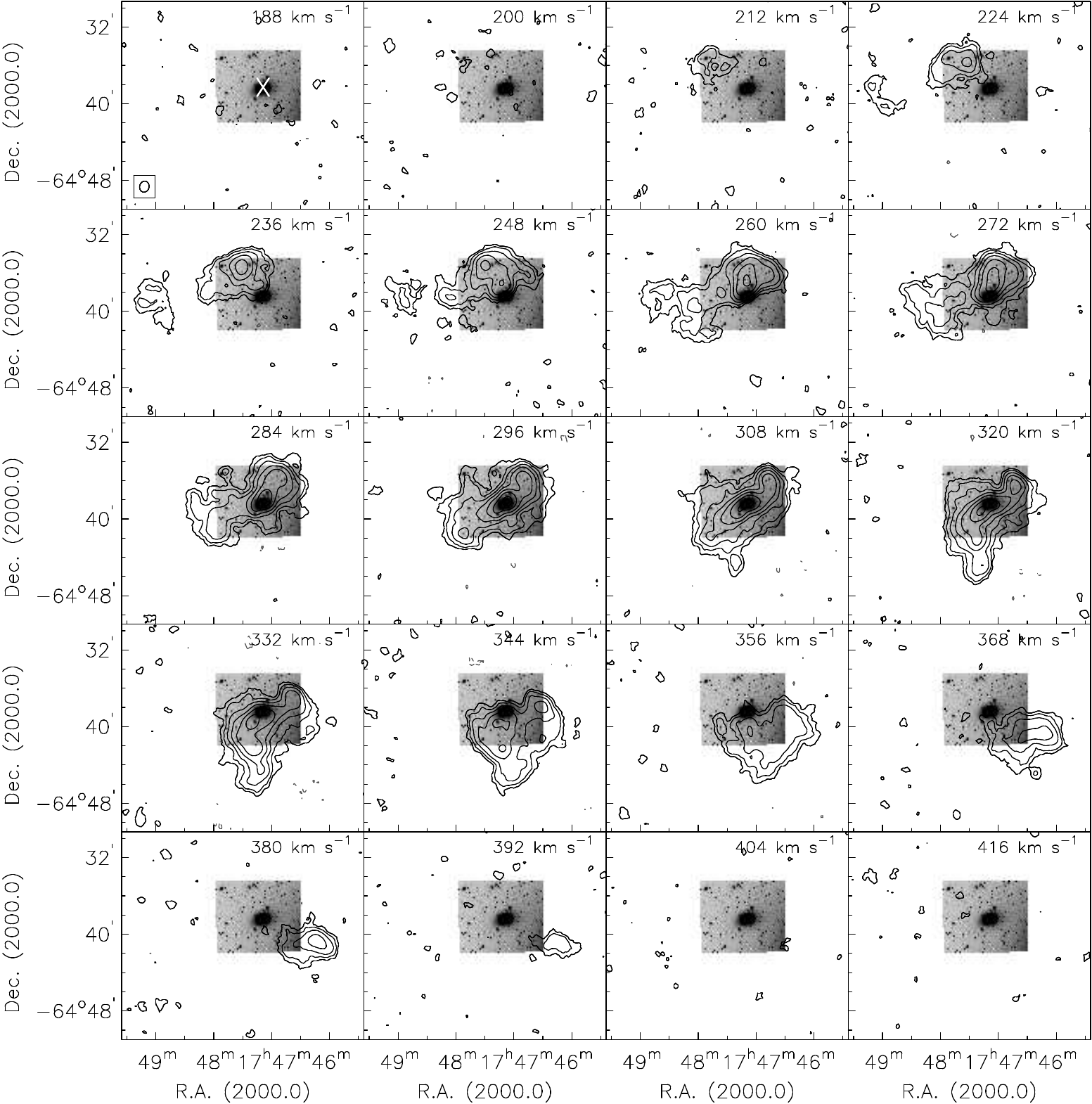}
\caption[Low-resolution \HI\ channel maps of IC\,4662 (contours) superposed on
  our \emph{R}-band image.]{Low-resolution \HI\ channel maps of IC\,4662
  (contours) superposed on our \emph{R}-band image. For display purposes we
  show the maps with a velocity resolution of 12\skms. The original channel
  spacing is 4\skms.The noise is about 0.9\,mJy\,beam$^{-1}$. Contours are
  drawn at $-$2.5 ($-$3$\sigma$), 2.5 (3$\sigma$), 5, 10, 20, 40, 80, and
  160\,mJy\,beam$^{-1}$. The synthesised beam of 70\arcsec$\times$62\arcsec\
  (natural weighting, excluding CA06) is displayed in the lower left corner of
  the first channel map. The optical centre of the galaxy is marked by a white
  cross in the same channel map. The corresponding heliocentric velocities are
  indicated in the upper right corner of each channel. 2\arcmin\ correspond to
  about 1.4\,kpc.}
\label{IC4662chan}
\end{figure*}

\begin{figure*}
\centering
\includegraphics[width=\textwidth]{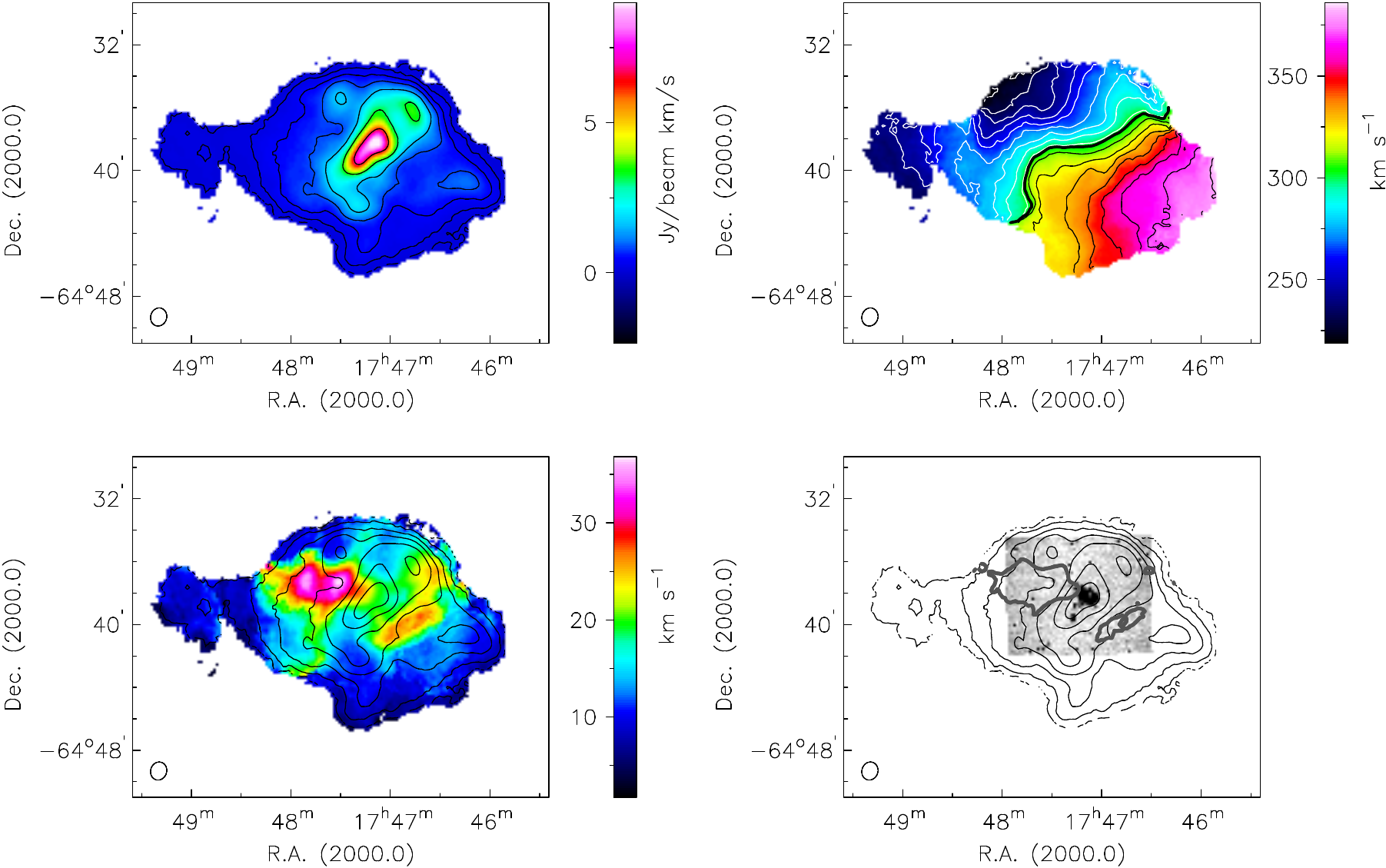}
\caption[\HI\ moment maps of IC\,4662.]{\HI\ moment maps of IC\,4662,
combining three arrays, but excluding baselines to CA06, which leads to a
synthesised beam of $70\arcsec \times 62\arcsec$ (natural weighting). {\bf Top
  left:} the \HI\ intensity distribution. Contours are drawn at 0.06
(3$\sigma$), 0.16, 0.4, 0.8, 1.6, 3.2, and 6.4\,Jy\,beam$^{-1}$\skms\ where
1\,Jy\,beam$^{-1}$ corresponds to a column density of $\rm 2.55\times
10^{20}\,atoms\,cm^{-2}$. {\bf Top right:} the \HI\ intensity-weighted mean
velocity field. Contour levels range from 220 to 380\skms\ in steps of
10\skms. The systemic velocity of 302\skms, estimated from the global \HI\
profile, is marked in bold. {\bf Bottom left:} the \HI\ velocity
dispersion. Overlaid are the same \HI\ intensity contours as mentioned
above. {\bf Bottom right:} continuum-subtracted \Ha\ image. The spatial
resolution was smoothed to fit the resolution of the \HI\ data. Overlaid in
black are again the \HI\ intensity contours. Overlaid in grey are the \HI\
velocity dispersion contours at 25\skms.
}
\label{IC4662mom}
\end{figure*}

The channel maps of the low-resolution \HI\ data (see Fig.~\ref{IC4662chan})
show that the emission is
distributed over a velocity range from roughly 200 to 400\skms, i.e., over
200\skms. The neutral gas seems to consist of two systems, an inner system
with a position angle of about 135\degr\ and an outer part with a position
angle of roughly 45\degr, i.e., perpendicular to the inner system. The
integrated \HI\ intensity map (upper left panel of Fig.~\ref{IC4662mom})
confirms this feature: it shows an inner high column density system that is
perpendicular to the outer low column density system. As displayed on the
lower right panel of Fig.~\ref{IC4662mom}, the inner system coincides with the
optical extent of IC\,4662 including the southern \HII\ region. We can see an
additional \HI\ maximum to the north-west of the galaxy, which has no optical
counterpart. The \HI\ outer diameter is given in
Table~\ref{TabHIproperties}. Note that the neutral gas is extended by a factor
of about six in comparison to the ionised gas, which is among the largest
ratios of \HI\ over \Ha ever detected. We measured the flux density to be $F_{\rm HI}=123$\,Jy\skms, from which we derived a total \HI\ mass of $M_{\rm
  HI}=1.7\times 10^8$\Msun\ (see Table~\ref{TabHIproperties}). This is
within the errors of the HIPASS value of $F_{\rm HI}=(130.0\pm12.0)$\,Jy\skms\
(see Table~\ref{basics}), which means that our interferometric observations
recover most of the emission detected on single-dish data of IC\,4662.

\begin{figure*}
\centering
\includegraphics[width=\textwidth]{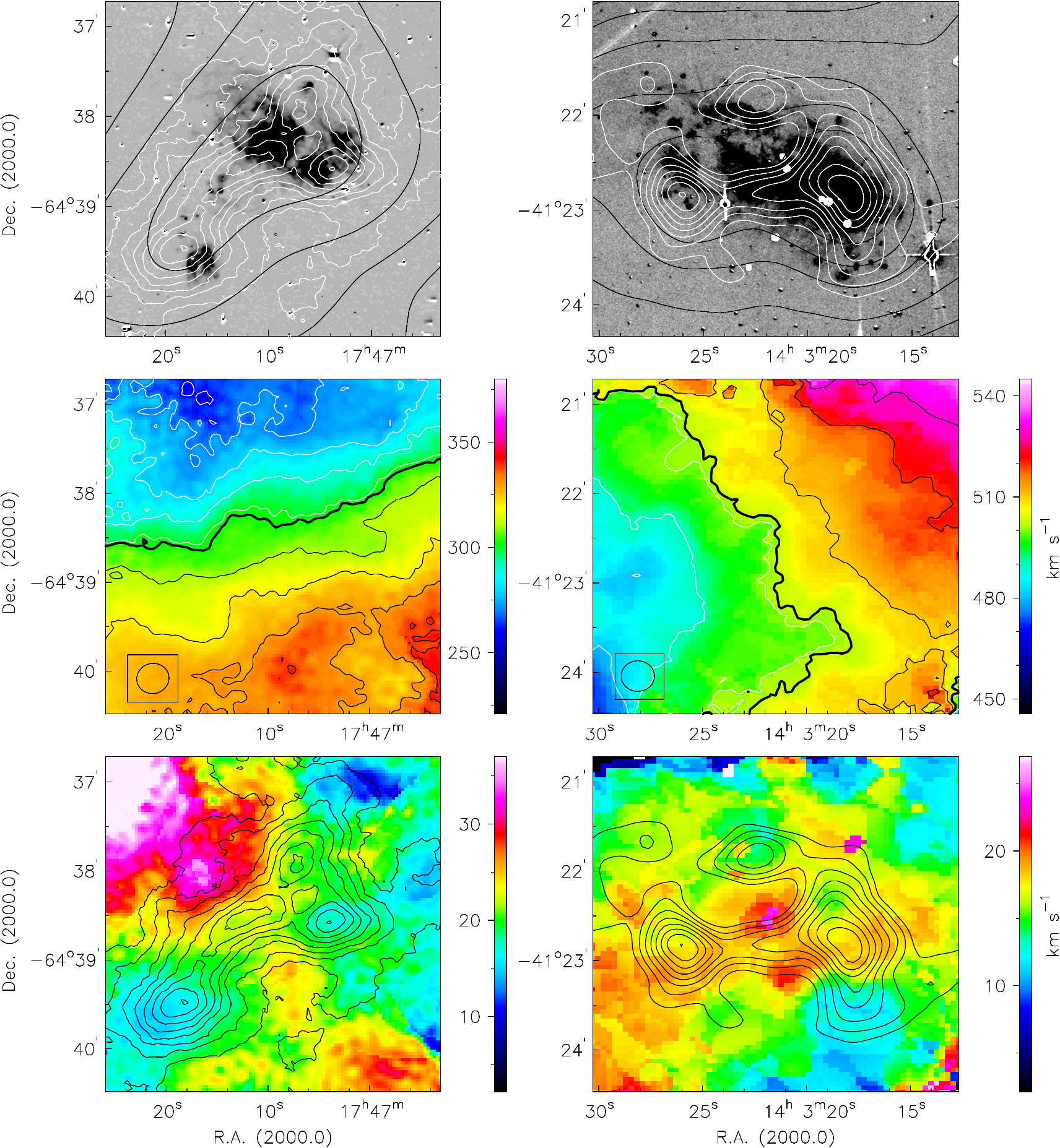}
\caption[]{High-resolution \HI\ moment maps of IC\,4662 (left) and NGC\,5408
  (right). {\bf Upper row:} a comparison of the \Ha\ and \HI\ line emission in
  the inner discs of both galaxies. The
  continuum-subtracted \Ha\ emission is shown as grey-scale, while the black
  and white contours indicate the \HI\ emission at low and high angular
  resolution respectively. The high-resolution contours of IC\,4662 are drawn
  at 0.2, 0.4, 0.6, 0.8, 1.0, 1.2, 1.4, and 1.6\,Jy\,beam$^{-1}$\skms\ where
  one beam corresponds to $21\arcsec \times 20\arcsec$; the ones of NGC\,5408
  at 1.2, 1.6, 2.0, 2.4, 2.8, 3.2, 3.6, and 4\,Jy\,beam$^{-1}$\skms\ where one
  beam corresponds to $20\arcsec \times 20\arcsec$. {\bf Middle row:} \HI\
  velocity fields. Contour levels are given in steps of 10\skms\ and range
  from 260 to 340\skms\ (IC\,4662) and from 480 to 530\skms\ (NGC\,5408). The
  systemic velocities (302\skms\ and 502\skms\ respectively) are marked in
  bold. {\bf Lower row:} \HI\ velocity dispersion maps. The high-resolution
  \HI\ intensity contours are overlaid as in the top panels.}
\label{momhres}
\end{figure*}

We also show the \HI\ intensity distribution at higher spatial resolution
(21\arcsec\,$\times$\,20\arcsec) for a better comparison with the optical
data. In the upper left panel of Fig.~\ref{momhres}, the high-resolution \HI\
intensity contours are overlaid in white onto the continuum-subtracted \Ha\
image. Additionally, the innermost contours of the low-resolution \HI\
intensity distribution are displayed in black. As the high-resolution contours
show, the central elongated maximum is split into several smaller, point-like
maxima. Interestingly, all these maxima are clearly offset from the most
prominent \HII\ regions.
\begin{table}
\caption{ATCA \HI\ properties.}
\label{TabHIproperties}
$$ 
\begin{tabular}{lcccc}
\hline
  \hline
Optical name         & IC\,4662  & NGC\,5408 \\
HIPASS name          & J1747--64 & J1403--41 \\
\hline
dynamic centre:  \\
~~$\alpha$(J2000)   & ... & 14$\rm ^h$ 03$\rm ^m$ 21.9$\rm ^s$ \\
~~$\delta$(J2000)    & ... & -41\degr\ 22\arcmin\ 03\arcsec \\
\vsys\ [\kms]  & 302 & 502\\
$i$ [\degr]            & ... & 62 \\
$PA$ [\degr]           & ... & 302 \\
$F_{\rm HI}$ [Jy\skms]       & 123 & 63\\
$M_{\rm HI}$ [10$^8$\,\Msun]    & 1.7 & 3.4\\
\HI\ diameter [\arcmin]        & 15.0 $\times$ 11.5 & 10.0 $\times$ 6.3 \\
\HI\ diameter [kpc]          & 10.6 $\times$ 8.2 & 14.0 $\times$ 8.8\\
 \HI\ / opt. ratio   & 6 & 4\\
\vrot\ [\kms]  & ... & 53 \\
$<\sigma>$ [\kms]    & 11/20 & 17 \\
$\sigma_{\rm Peak}$ [\kms] & 37 & 27 \\
$r_{\rm HI,max}$ [kpc] & 6 & 9.5\\
$M_{\rm dyn}$ [$\rm 10^9$\,\Msun] & ... & 6.2\\
\hline
  \hline
\end{tabular}
$$
\flushleft\footnotesize{
Note: As the velocity field of IC\,4662 is very distorted, we did not
perform a tilted-ring analysis and therefore could not calculate any of the
kinematic parameters. The systemic velocity was estimated from the global \HI\
profile of IC\,4662. The lower value of $<\sigma>$ refers to the outer system
of IC\,4662, the higher value to the inner, perpendicular system.
}
\end{table}
\subsection{NGC\,5408}
\label{SectN5408morpho}
Like IC\,4662, NGC\,5408 reveals a slightly elongated stellar disc with a
position angle of 62\degr, which leads to a box-like appearance in the optical
(see Fig.~\ref{Rband}, lower left panel). The main star-forming knots are
located along a chain running from the centre to the main star formation area
in the south-west. In \Ha, the
galaxy is dominated by a bright complex of \HII\ regions in the south-west
(Fig.~\ref{Rband}, lower right panel) that coincides with the brightest star
clusters. In the eastern part, the ionised gas becomes more patchy with a few
holes in the distribution and several filaments emanating from the
disc. Indeed, it seems that the eastern part of the galaxy ends in a
superbubble as suggested by the shell-like structures visible in the north- and
south-east and the big hole in between. Diffuse gas can also be seen across
the southern part of the galaxy.

\begin{figure*}
\centering
\includegraphics[width=\textwidth]{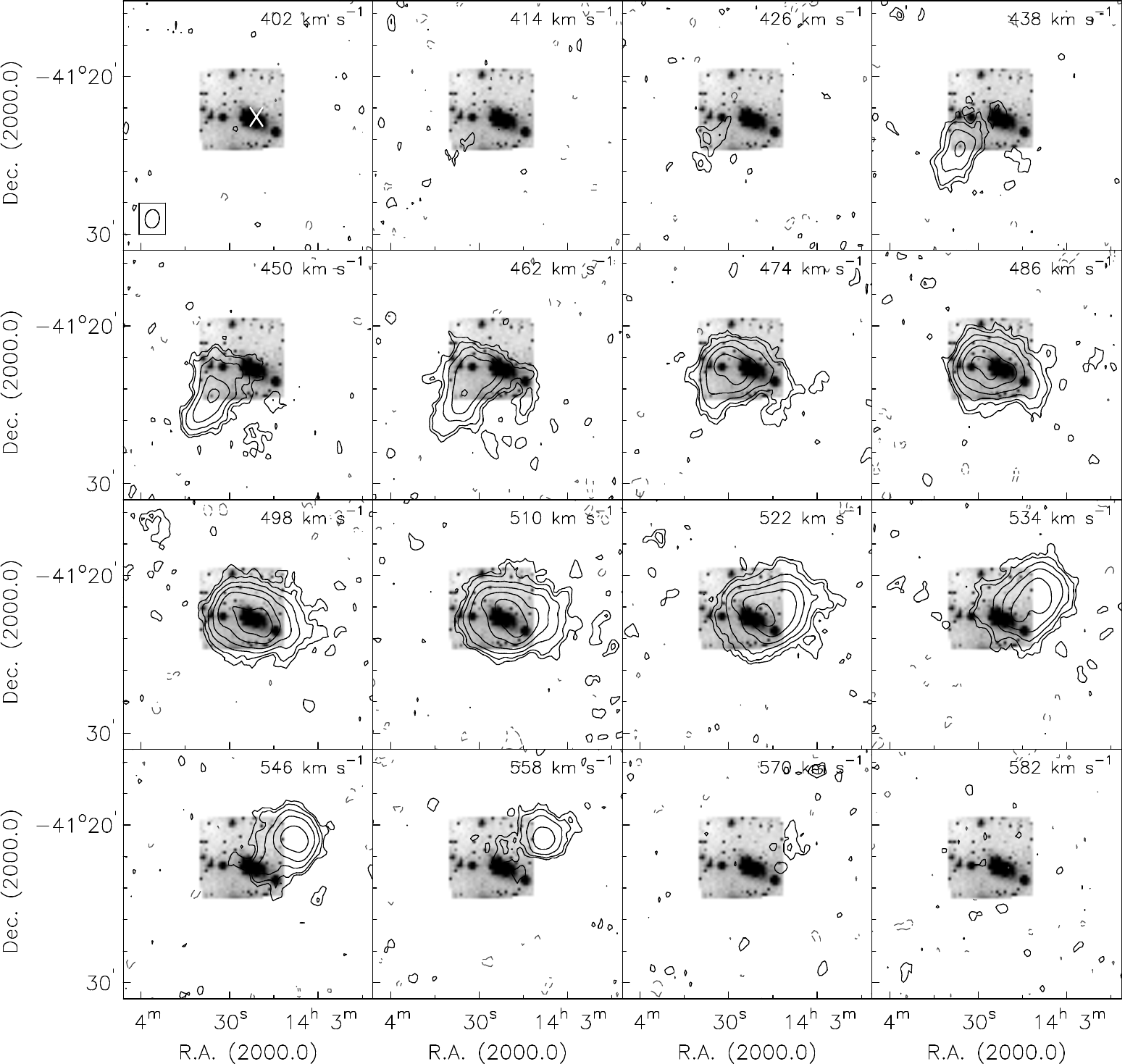}
\caption[Low-resolution \HI\ channel maps of NGC\,5408 (contours) superposed
  on the \emph{R}-band image.]{Low-resolution \HI\ channel maps of NGC\,5408 (contours) superposed on
the \emph{R}-band image. For display purposes we show the maps with a velocity
resolution of {\bf 12}\skms. The original channel spacing is 4\skms. The noise is about 1\,mJy\,beam$^{-1}$. Contours are
drawn at -3 ($-$3$\sigma$), 3 (3$\sigma$), 5, 10, 20, 40, 80, and
160\,mJy\,beam$^{-1}$. Otherwise the same as in Fig.~\ref{IC4662chan}. The
synthesised beam is 71\arcsec $\times$\,54\arcsec\ (robust weighting,
excluding CA06). 2\arcmin\ correspond to 2.8\,kpc.}
\label{N5408chan}
\end{figure*}

\begin{figure*}
\centering
\includegraphics[width=\textwidth]{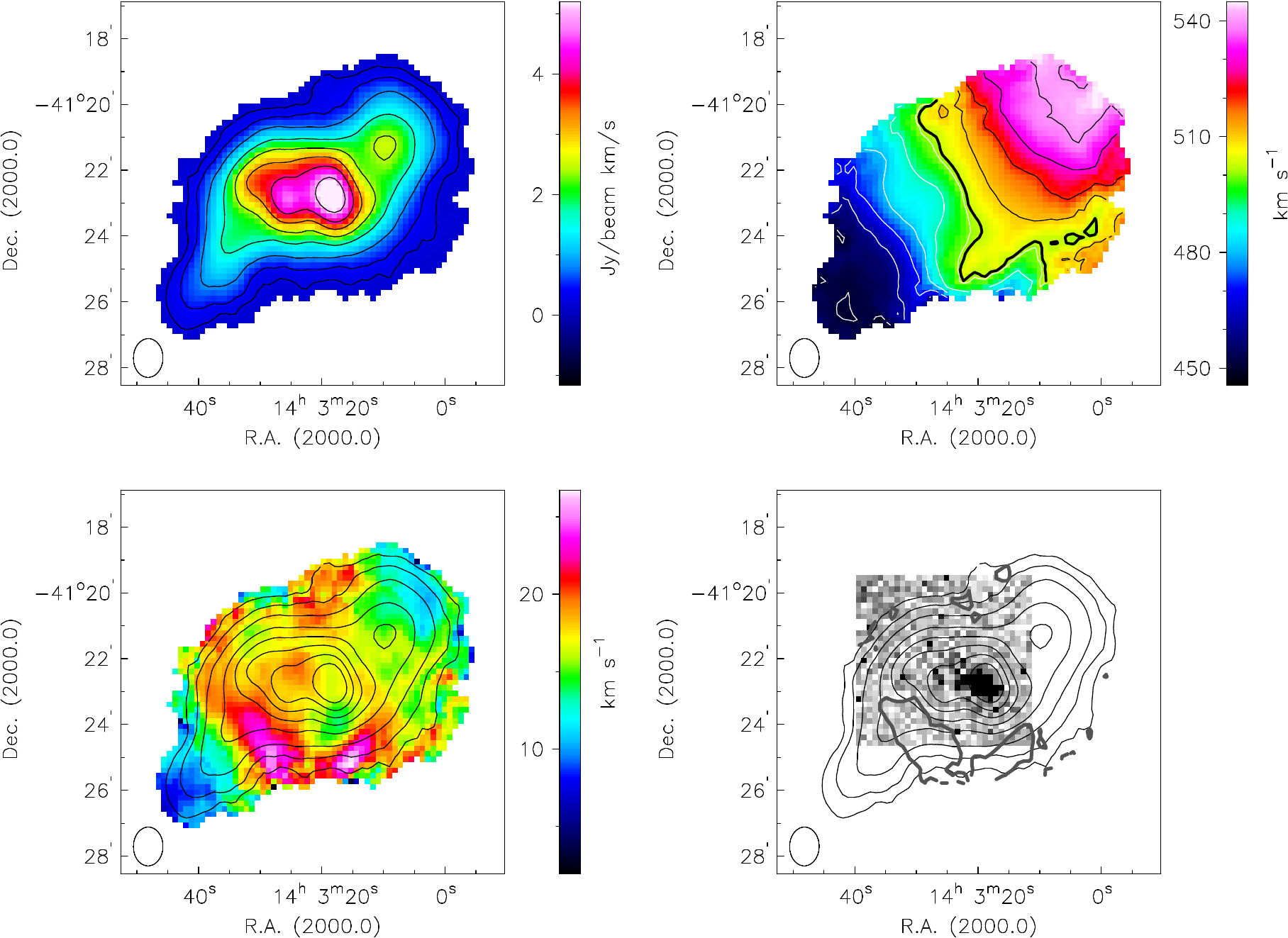}
\caption[\HI\ moment maps of NGC\,5408.]{\HI\ moment maps of NGC\,5408,
combining three arrays, but excluding all baselines to CA06, which leads to a
synthesised beam of $71\arcsec \times 54\arcsec$ (robust weighting). {\bf Top
  left:} the \HI\ intensity distribution. Contours are drawn at 0.2
(3$\sigma$), 0.4, 0.8, 1.2, 2.0, 3.2, and 4.0\,Jy\,beam$^{-1}$\skms\ where
1\,Jy\,beam$^{-1}$ correspond to a column density of $\rm 2.89\times
10^{20}\,atoms\,cm^{-2}$. {\bf Top right:} the 
\HI\ intensity-weighted mean velocity field. Contour levels range from 450 to
550\skms\ in steps of 10\skms. The systemic velocity of 502\skms, as
calculated from fitting tilted-rings to the velocity field, is marked in
bold. {\bf Bottom left:} the \HI\ velocity dispersion, overlaid are the same
\HI\ intensity contours as mentioned above. {\bf Bottom right:}
continuum-subtracted \Ha\ image. The spatial resolution was smoothed to fit
the resolution of the \HI\ data. Overlaid in black are again the \HI\
intensity contours. Overlaid in grey are the \HI\ velocity dispersion contours
at 20\skms.}
\label{N5408mom}
\end{figure*}

Figure~\ref{N5408chan} shows the low-resolution \HI\ channel maps of
NGC\,5408. We detected emission over 130\skms, from 430 to 560\skms. In
  the inner parts the emission is elongated along the east-west axis
  (position angle of 90\degr), whereas in the outer parts it is more extended
  to the north-west and south-east (position angle of 135\degr). The integrated \HI\ intensity distribution looks
fairly symmetric and reveals two point-like maxima in the centre (see
Fig.~\ref{N5408chan} and upper left panel of Fig.~\ref{N5408mom}). A
comparison with the ionised gas distribution shows that the western \HI\
intensity maximum coincides with the centre of the optical emission in
NGC\,5408 (see Fig.~\ref{N5408mom}, lower right panel). On larger scales, the
distribution of the neutral gas seems to be warped in the outer parts, which
explains the two orientations seen in the channel maps
(Fig.~\ref{N5408chan}). In comparison to the ionised gas, the neutral gas component is four times more extended, which is
again an unusually high ratio (for the \HI\ diameter see
Table~\ref{TabHIproperties}). We measured the flux density to be $F_{\rm
  HI}=63$\,Jy\skms, from which we derived a total \HI\ mass of $M_{\rm
  HI}=3.4\times 10^8$\Msun\ (Table~\ref{TabHIproperties}). This is again
  within the errors of the HIPASS value of $F_{\rm HI}=(61.5\pm6.7)$\,Jy\skms.

Again, we overlaid the high-resolution \HI\ intensity contours
(20\arcsec\,$\times$\,20\arcsec) in white over the continuum-subtracted \Ha\
image (Fig.~\ref{momhres}, upper right panel). The distribution shows two
prominent maxima which coincide with the point-like maxima detected on the
low-resolution \HI\ map (black contours). The western extended maximum
coincides with the main \HII\ region complex in NGC\,5408, although it is not
fully centred on the brightest \HII\ region. It is connected via
a bridge with the eastern maximum which is clearly offset from the southern
shell-like structure mentioned above. Additionally, the high-resolution \HI\
intensity distribution reveals a third, weaker maximum in the north, close to,
but offset from the northern shell-like structure. In agreement with the
distribution of the ionised gas, the \HI\ column density between the
shell-like structures is very low. This confirms our findings in the optical
data, i.e., the existence of a superbubble in the eastern part of NGC\,5408. 
\section{Kinematic analysis}
The comparison of the optical and \HI\ morphology in IC\,4662 and NGC\,5408
showed that on low-resolution scales, the optical extent coincides very well
with the central \HI\ maximum. However, the high-resolution \HI\ maps reveal
that the peaks in the neutral gas distribution are typically offset from the
\HII\ regions. On larger scales, the optical appearance of both galaxies is
defined by an alignment which is perpendicular to the extended \HI\
distribution (IC\,4662) or differs by at least 45\degr\ (NGC\,5408).

In this section, we analyse the kinematics of the two galaxies in order to
study the properties of both their neutral and ionised gas components. All
given velocities are heliocentric velocities measured along the line of sight.
\subsection{The distorted \HI\ kinematics of IC\,4662}
\label{I4662HIkinematics}
The \HI\ velocity field of IC\,4662 (Fig.~\ref{IC4662mom}, upper right
panel; see also Fig.~\ref{momhres}, middle row, left panel) is very disturbed. The
overall velocity gradient runs from the north-east with velocities of
220\skms\ to the south-west with velocities of 380\skms. The direction of the
rotation of the neutral gas seems to change from the western part of the
galaxy to the eastern part, causing a sudden change of the position angle of
almost 90\degr. This change is possibly connected with the two perpendicular
systems as discussed in Sect.~\ref{SectI4662morpho}. The \HI\ velocity
dispersion map (lower left panel of Fig.~\ref{IC4662mom}, overlaid are the
\HI\ intensity contours; see also Fig.~\ref{momhres}, lower left panel) shows
a dispersion of 20\skms\ in the inner and 11\skms\ in the outer parts, except
for two maxima with almost 40\skms\ east of the extended \HI\ intensity
maximum.

\begin{figure*}
\centering
\includegraphics[width=\textwidth]{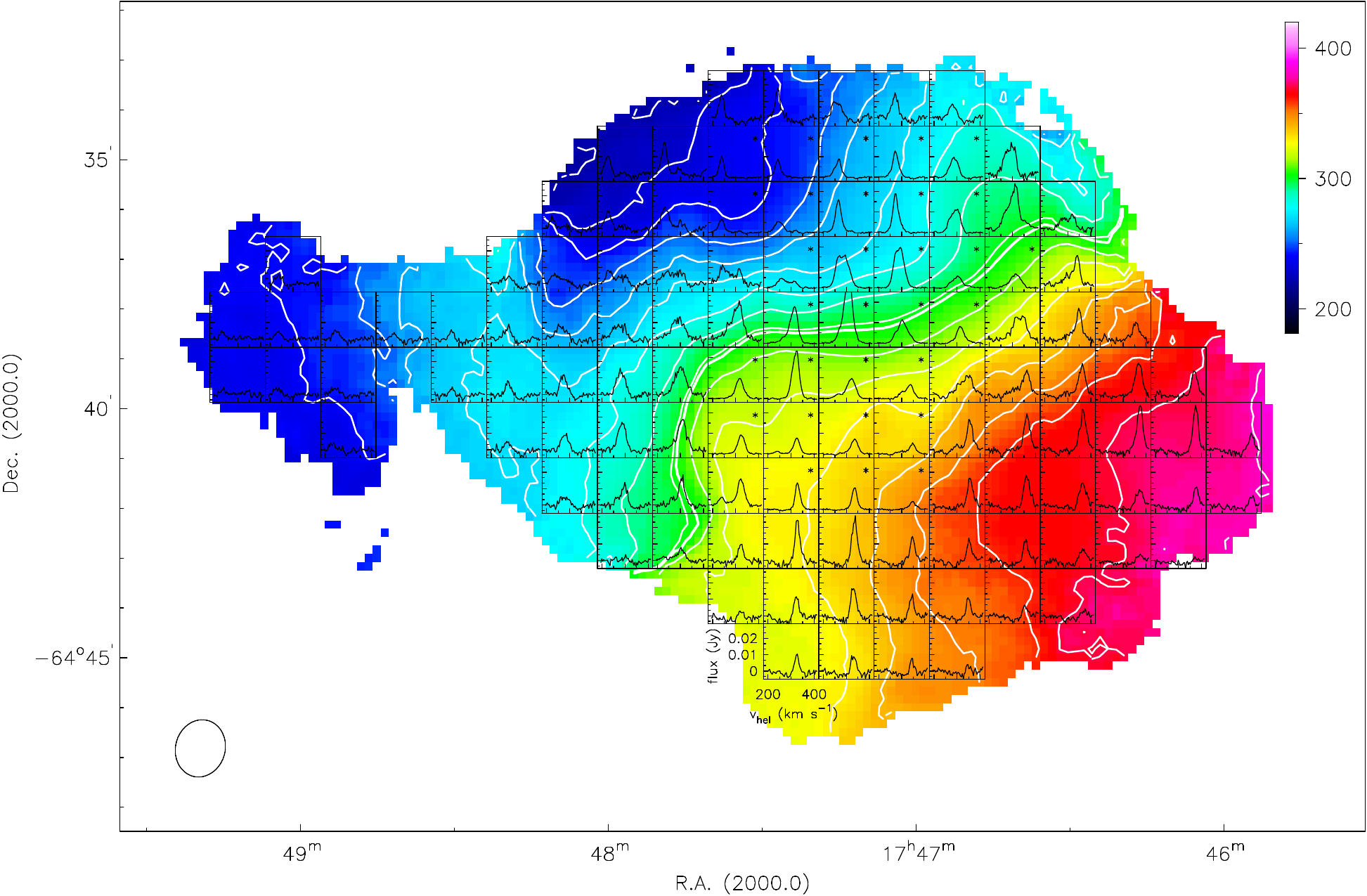}
\caption{\HI\ velocity profiles of IC\,4662. We show the low-resolution \HI\
  velocity field (see Fig.~\ref{IC4662mom}, upper right panel) with the \HI\
  profiles overlaid. Each box represents the sum of all spectra within roughly
  one beam size (70\arcsec\,$\times$\,70\arcsec). Stars denote spectra with a flux scale which is four times
  larger than that of the other spectra.
}
\label{IC4662profiles}
\end{figure*}

The large-scale twist in the velocity field makes it impossible to derive a
rotation curve for IC\,4662. We extracted spectra from the \HI\ data cube to
get a better understanding of the kinematics of the neutral
gas. They are shown in Fig.~\ref{IC4662profiles}: each box represents the sum
of all spectra within 70\arcsec\,$\times$\,70\arcsec\ (corresponding to one
beam size). Stars denote spectra with a flux scale which is four times
larger than that of the other spectra. It is obvious that across the whole
distorted area, the \HI\ line is often split into two, sometimes clearly
separated components. The difference of about 70\skms\ is always the same,
although the absolute velocities vary according to the rotation of IC\,4662.

As already mentioned in Sect.~\ref{sectradiored}, we created the velocity
field using the intensity-weighted mean (iwm). However, as soon as the velocity
distribution of the \HI\ emission is not symmetric, the velocity values are
biased to the longest tail of the velocity profiles
\citep{deBlok2008}. Therefore, one might argue that the iwm
velocity field does not represent the true velocities. A recently more popular
approach is to fit the line profiles with Gauss-Hermite polynomials, which
allows to define the peak velocities more accurately. This becomes important
when components overlap. In the IC\,4662 spectra, the two components are
clearly separated with one of them dominating the \HI\ profile. Hence, the
difference between the iwm and the Hermite velocity field should be
negligible. Nevertheless, we created the Hermite velocity field by fitting
Gauss-Hermite h3 polynomials to all line profiles using the GIPSY task
\emph{xgaufit} (Fig.~\ref{IC4662iwm-herm}, middle
panel). The right panel of Fig.~\ref{IC4662iwm-herm} shows the residuals after
subtracting the Hermite velocity field from the iwm. As we expected, the
residuals are mostly close to zero, except for an area in the north-east. A
look at the spectra (Fig.~\ref{IC4662profiles}) reveals that the S/N in this
area is quite low, which can often not be handled properly by the GIPSY
\emph{moments} task which we used to create the iwm velocity field. As the
large-scale distortion is still very pronounced in the Hermite velocity field,
we think that it is not a consequence of using the 'wrong' method to create
the velocity field, but that it is a real feature in IC\,4662.
\subsection{The \HI\ velocity field of NGC\,5408}
NGC\,5408 shows a symmetric \HI\ velocity field (Fig.~\ref{N5408mom},
upper right panel; see also Fig.~\ref{momhres}, middle row, right panel) with a velocity
gradient running from the south-east with velocities of 450\skms\ to the
north-west with velocities of 550\skms. The gradient is smooth, even at both
ends of the disc where the \HI\ distribution looks warped. It even seems that
it better follows the large scale structure than the inner east-west
alignment. A local, quite prominent distortion of the velocity field can be
seen south of the \HI\ intensity maxima. The \HI\ intensity distribution,
however, appears to be smooth and undisturbed (see upper left panel of
Fig.~\ref{N5408mom}). The \HI\ velocity dispersion (lower left panel of
Fig.~\ref{N5408mom}, overlaid are the \HI\ intensity contours; see also
Fig.~\ref{momhres}, lower right panel) shows two peaks at 25\skms, which are
located on either side of the spur in the velocity field.

\begin{figure*}
\centering
\includegraphics[width=.7\textwidth]{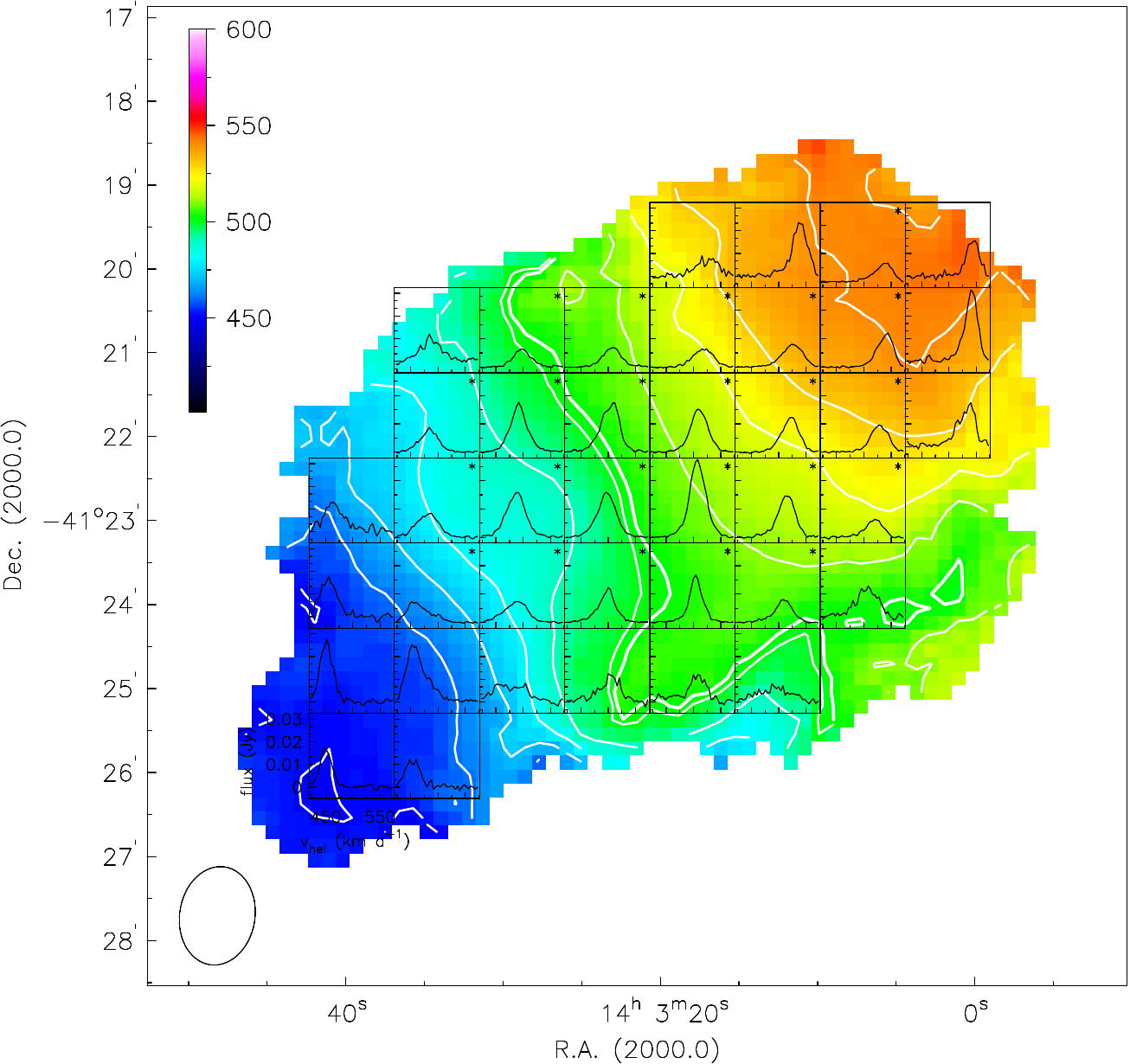}
\caption{\HI\ velocity profiles of NGC\,5408. We show the low-resolution \HI\
  velocity field (see Fig.~\ref{N5408mom}, upper right panel) with the \HI\
  profiles overlaid. Each box represents the sum of all spectra within roughly
  one beam size (60\arcsec\,$\times$\,60\arcsec). Stars denote spectra with
  a flux scale which is four times larger than that of the other spectra.
}
\label{N5408profiles}
\end{figure*}

Again, we extracted the spectra from the \HI\ data cube in order to analyse
the profiles (see Fig.~\ref{N5408profiles}). Each box equals a
beam size of 60\arcsec\,$\times$\,60\arcsec. Stars denote spectra with a
flux scale which is four times larger than that of the other spectra. In
general, the \HI\ profiles look quite simple. The lines are broad with
\emph{FWHM} of 35 to 50\skms, but often also
symmetric with no second component clearly standing out. The high velocity
dispersion in the south coincides with an area of very low S/N, which is
probably again due to the inability of the \emph{moments} task to properly
handle noisy spectra. However, at the eastern edge we detected two
components which are separated by about 40\skms. 

We tested the quality of the iwm velocity field by subtracting the Hermite
velocity field: the residuals are very small except for the area of high
velocity dispersion. As in the case of IC\,4662, this can be explained by the
low S/N and the line-splitting. For a direct comparison we refer the reader to
\citet{vanEymeren2009b} where the \HI\ kinematics of several dwarf galaxies
(among them NGC\,5408) have been analysed by using the Hermite velocity fields.
\subsection{The \HI\ rotation curve of NGC\,5408}
\label{ICNGCrot}
\begin{figure*}
\centering
\includegraphics[width=\textwidth]{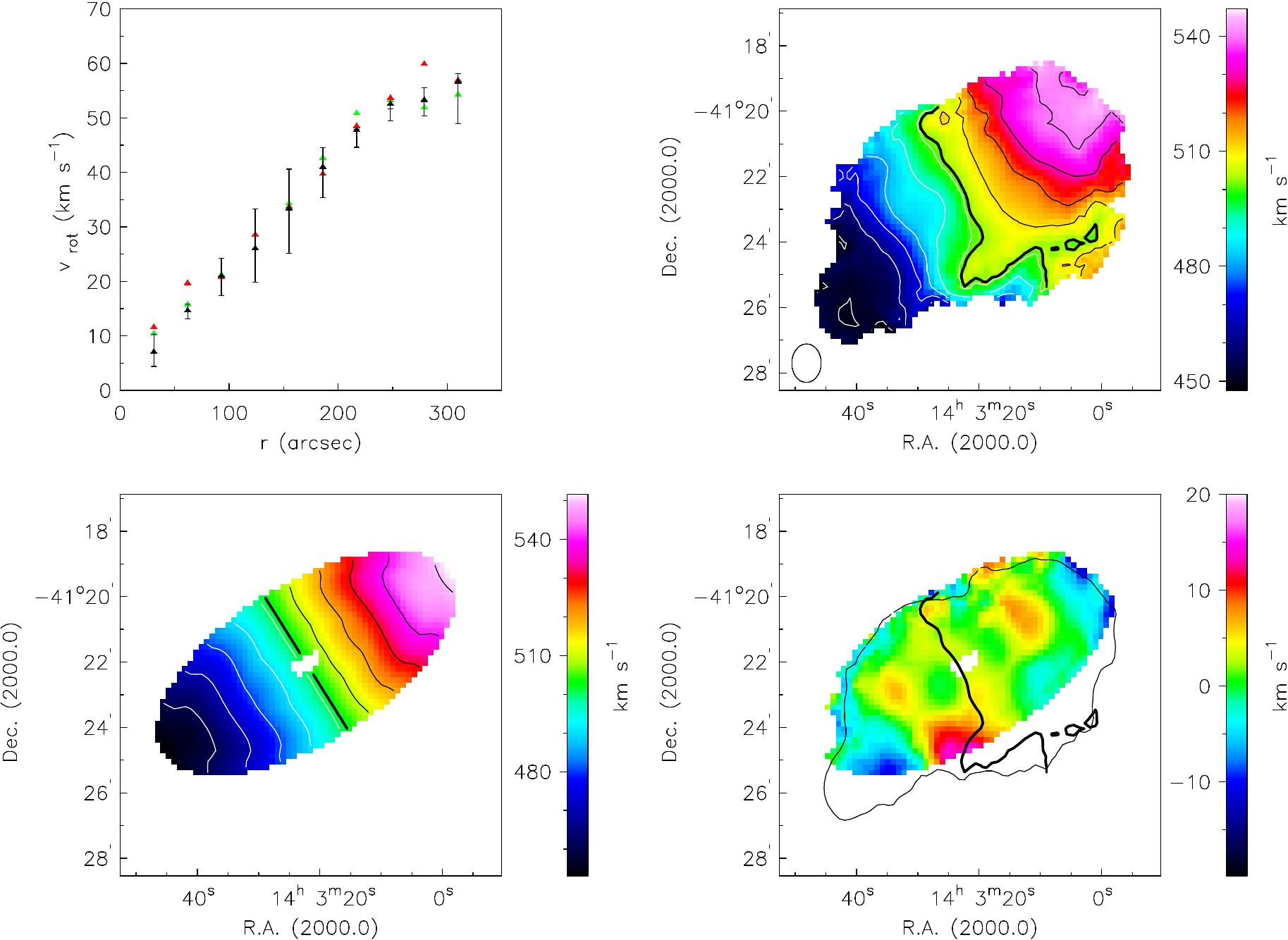}
\caption[The kinematics of NGC\,5408.]{The \HI\ rotation curve of NGC\,5408
  created by fitting a tilted-ring model to the observed velocity field. {\bf
  Top left:} different approaches for deriving the rotation curve. The
  black triangles represent the iterative approach, the error bars indicate
  receding and approaching sides. The green (light grey) curve was derived by
  taking the initial estimates and keeping them fixed, the red (dark grey)
  curve by taking the best-fitting parameters and leaving them free. {\bf Top
  right:} the observed \HI\ velocity field (see Fig.~\ref{N5408mom}). {\bf
  Bottom left:} the model velocity field created by taking the best-fitting
  parameters. {\bf Bottom right:} the residual map after subtracting the model
  from the original velocity field. The 3$\sigma$ contour from the
  low-resolution \HI\ intensity map and the systemic velocity are overplotted
  in black.}
\label{N5408rot}
\end{figure*}

In order to make predictions about the fate of the gas, we need to know some
of the kinematic parameters like the inclination and rotation velocity of
NGC\,5408. Therefore, we derived a rotation curve by fitting a tilted-ring
model to the observed velocity field \citep[GIPSY task
  \emph{rotcur},][]{Begeman1989}. Initial estimates for the relevant
parameters (systemic velocity, the coordinates of the dynamic centre, the
inclination, and the position angle) were obtained by interactively fitting
ellipses to the \HI\ intensity distribution with the GIPSY task
\emph{ellfit}. The width of the rings was chosen to be half the spatial
resolution, i.e., 31\arcsec. In order to get the most precise values, three
different approaches were made by always combining receding and approaching
side. The upper left panel of Fig.~\ref{N5408rot} shows the resulting curves:
first, we kept the initial estimates fixed in order to derive the rotation
velocities (green (light grey) triangles). The best-fitting values for all
parameters were derived by allowing only one parameter to vary with radius
while keeping the remaining ones fixed. As no significant variation of any of
the parameters over radius was noticed, we derived an average value (given in
Table~\ref{TabHIproperties}). The black triangles represent the resulting
rotation velocities. The error bars indicate the values of the receding (top)
and approaching side (bottom) when treated separately. In order to reproduce
the result of the iterative approach, the so-derived parameters were all left
free (red (dark grey) triangles). As the filling factor of the rings drops
significantly beyond a radius of 320\arcsec, we did not fit the outer
parts. That means we also miss most of the distortion at the southern edge of
the galaxy.

NGC\,5408 has a slowly rising rotation curve coming to a plateau at a radius
of 250\arcsec. Receding and approaching side show a very similar kinematic
behaviour up to a radius of 320\arcsec, confirming the impression of an evenly
rotating galaxy. The initial estimates already defined the \HI\ kinematics
quite well as the green and black triangles are in very good agreement. The
'left-free' approach is also in good agreement with the other two approaches,
except for some scatter between 0\arcsec\ and 100\arcsec. This is still far
away from the disturbed region, which can therefore not explain the
scatter.

The iterative approach results in an inclination angle of 62\degr, a position
angle of 302\degr, and a systemic velocity of 502\skms\ (see also
Table~\ref{TabHIproperties}). The systemic velocity is slightly lower than the
HIPASS velocity of 506\skms\ \citep{Koribalski2004}. In
comparison to the parameters derived from optical data, the inclination of the
neutral gas is higher by 10\degr. The value of the position angle of the
neutral gas is dominated by the extended emission. As the optical disc is
differently aligned (see Sect.~\ref{SectN5408morpho}), the \HI\ position angle
of 302\degr\ differs significantly from the optical position angle of 60\degr. 

In order to prove the reliability of the calculated values, a model velocity
field was created by using the best-fitting parameters (Fig.~\ref{N5408rot},
lower left panel) and afterwards subtracted from the original velocity field
(upper right panel). The residual map is shown on the lower right panel of
Fig.~\ref{N5408rot} with the 3$\sigma$ contour from the low-resolution \HI\
intensity map and the systemic velocity contour overplotted. It is obvious
that the original velocity field is very well described by the derived
kinematic parameters. The residuals are generally at $\pm$\,5\skms, except for
one region in the south, which shows residuals of about 15\skms. This region
adjoins the already mentioned distortion in the velocity field and the high
velocity dispersion. We examine this issue in more detail in
Sect.~\ref{N5408longslit}.
\subsection{A detailed kinematic analysis of the \HI\ and \Ha\ line data}
\label{comp_spec}
We now want to study the \HI\ and \Ha\ line profiles of both galaxies in
  more detail and to compare the peak velocities of the neutral and ionised
  gas. Therefore, we performed a Gaussian decomposition by interactively
  fitting the \HI\ and \Ha\ line emission (IRAF task \emph{splot}). Only
  detections above a 3$\sigma$ threshold were considered.

\begin{figure*}
\centering
\includegraphics[width=\textwidth]{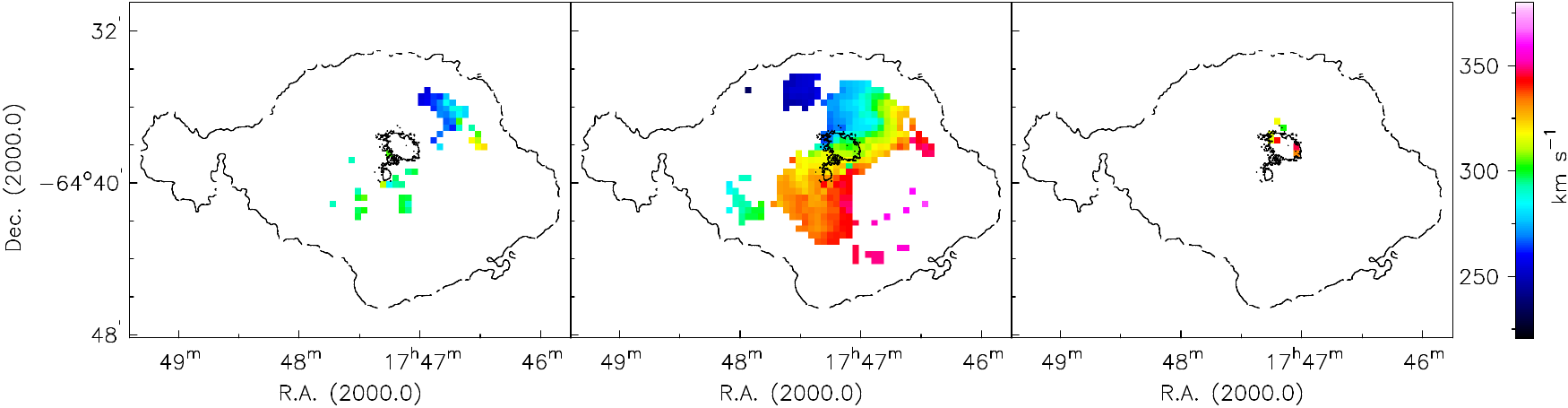}
\caption[Gaussian decomposition of the \HI\ in IC\,4662.]{Gaussian
decomposition of the \HI\ line profiles of IC\,4662, which were extracted from
the high-resolution cube. We show the blue-shifted (left panel), main (middle
panel) and red-shifted (right panel) components of the \HI\ velocities. The
3$\sigma$ \HI\ intensity contour from the low-resolution cube and the
3$\sigma$ \Ha\ intensity contour from the \Ha\ image (Fig.~\ref{Rband}) are
overlaid in black for better visualisation. We averaged the velocities over
20\arcsec\,$\times$\,20\arcsec, which roughly corresponds to one beam size.}
\label{ic4662.hi.decomp}
\end{figure*}

\begin{figure*}
\centering
\includegraphics[width=\textwidth]{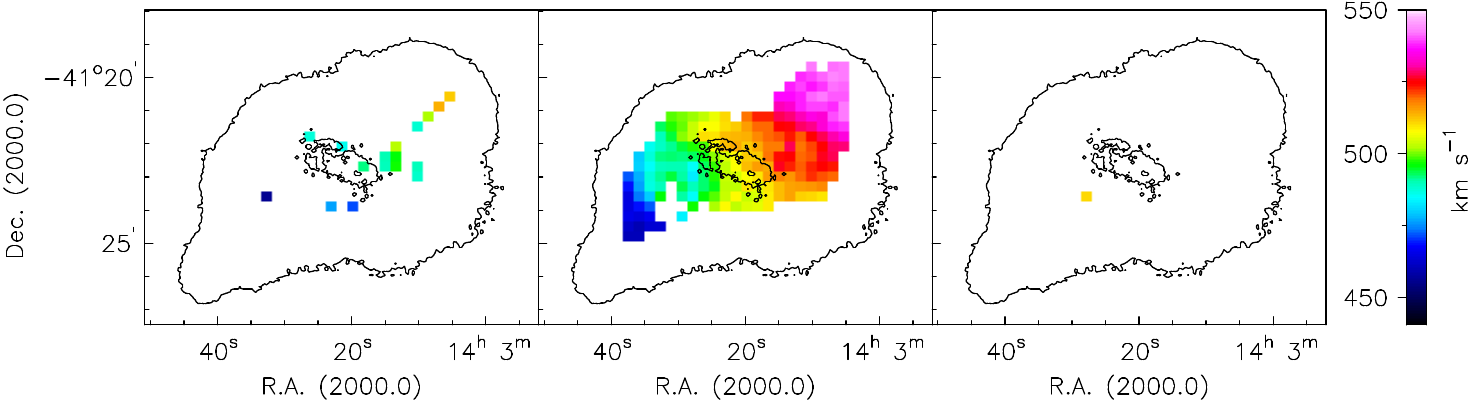}
\caption[Gaussian decomposition of the \HI\ in NGC\,5408.]{Gaussian
decomposition of the \HI\ line profiles of NGC\,5408. The same as in
Fig.~\ref{ic4662.hi.decomp}. We averaged the velocities over
18\arcsec\,$\times$\,18\arcsec, which roughly corresponds to one beam size.}
\label{n5408.hi.decomp.hres}
\end{figure*}

Figures~\ref{IC4662profiles} and \ref{N5408profiles} have already shown that
the \HI\ profiles of both galaxies are sometimes split into at least two
components. For a better comparison with the \Ha\ images, we used the
high-resolution \HI\ data cubes (see Sect.~\ref{sectradiored}) to perform the
Gaussian decomposition. As the extracted spectra are often quite noisy,
especially in the outer parts of both galaxies, only the central parts are
fitted. Figures~\ref{ic4662.hi.decomp} and \ref{n5408.hi.decomp.hres} show the
resulting maps: a map with the blue-shifted gas on the left, a map with the
main component (component of highest intensity) in the middle and a map with
red-shifted gas on the right. The velocities are averaged over
20\arcsec\,$\times$\,20\arcsec\ in case of IC\,4662 and
18\arcsec\,$\times$\,18\arcsec\ in case of NGC\,5408, which roughly
corresponds to one beam size. For better visualisation, we overlaid the
3$\sigma$ \HI\ intensity contour from the low-resolution data and the
3$\sigma$ \Ha\ intensity contour from the \Ha\ images (Fig.~\ref{Rband}).

\begin{figure*}
\centering
\includegraphics[width=\textwidth]{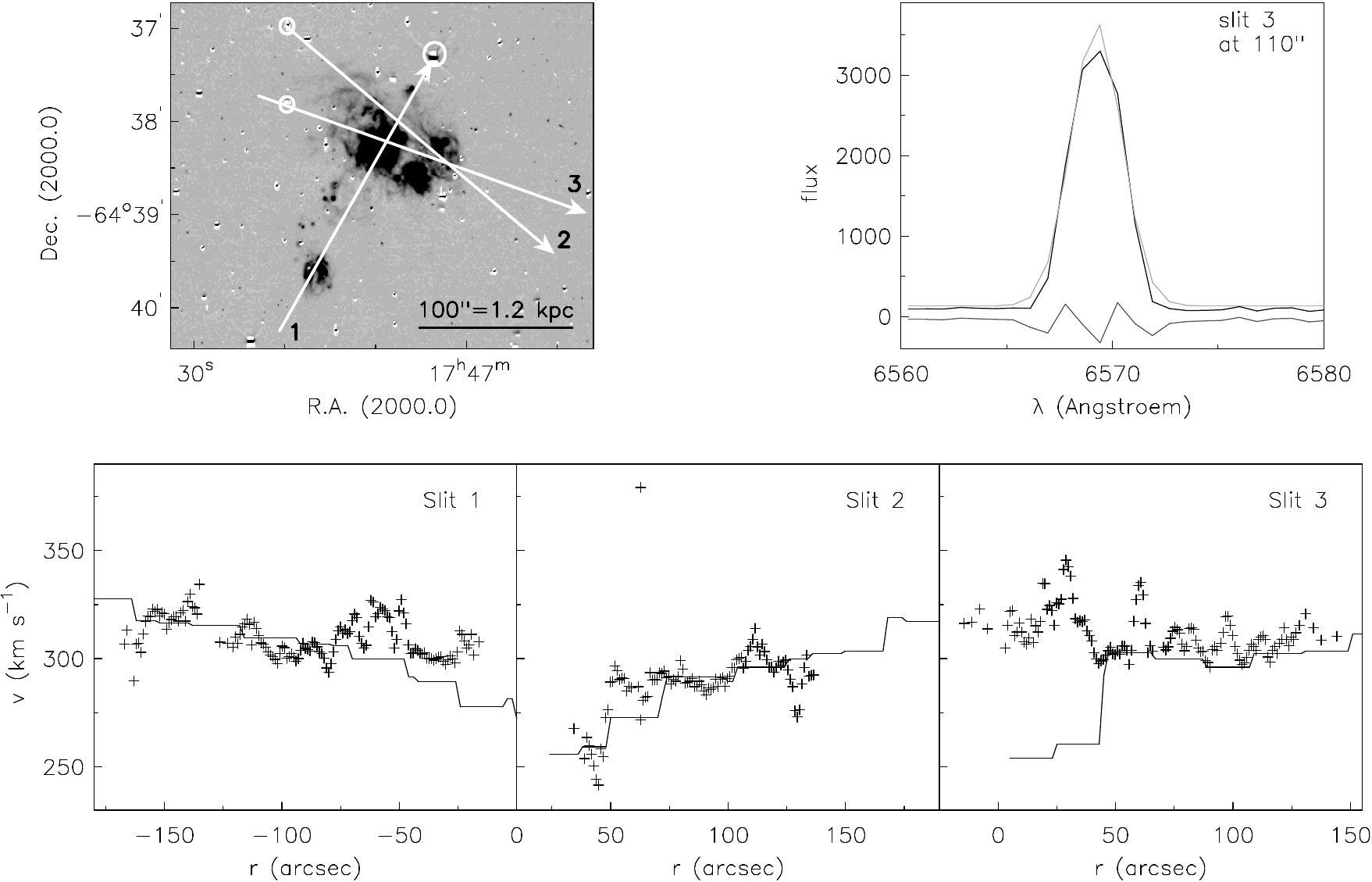}
\caption[Pv diagrams of IC\,4662.]{Gaussian decomposition of the \Ha\ line
    profiles of IC\,4662. {\bf Upper left
    panel:} the continuum-subtracted \Ha\ image. White arrows indicate the slit
positions, white circles mark position 0. {\bf Upper right panel:} an example
    spectrum (slit\,3 at $r=110$\arcsec) which shows the flux in arbitrary
    units \emph{vs.} wavelength. The black solid line indicates the original
    spectrum, the light grey line marks the Gaussian fit. The residuals
    after subtracting the Gaussian from the original profile are shown in dark
    grey. Note that the spectral resolution is about
    2.5\AA, which corresponds to 112\skms. {\bf Lower panels:} the
pv diagrams of the \Ha\ emission. The + symbols represent the \Ha\ velocities
obtained from the spectra, the solid line the \HI\ velocities obtained from
the high-resolution velocity map of the main component (see Fig.~\ref{ic4662.hi.decomp}, middle panel).}
\label{ic4662pv}
\end{figure*}

\begin{figure*}
\centering
\includegraphics[width=\textwidth]{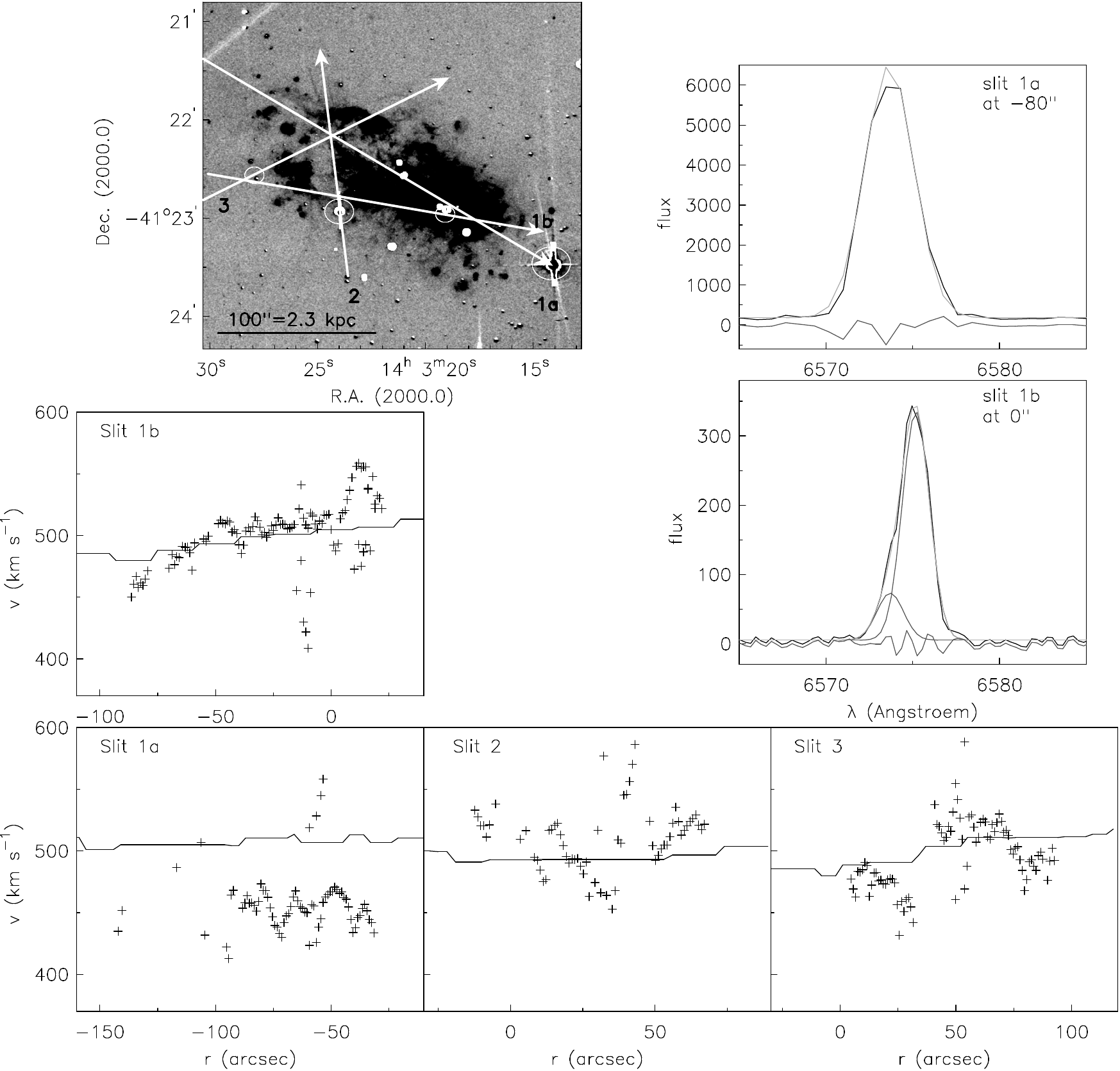}
\caption[Pv diagrams of NGC\,5408.]{Gaussian decomposition of the \Ha\ line
    profiles of NGC\,5408. {\bf Upper left
    panel:} the continuum-subtracted \Ha\ image. Again, all slit positions are
    indicated by white arrows. Position 0 is marked by white circles. {\bf
    Upper and middle right panels:} two example spectra (slit\,1a at
    $r=-80$\arcsec, slit\,1b at $r=0$\arcsec). Black solid lines indicate the
    original spectra, light grey lines mark the Gaussian fit in case of
    slit\,1a and the sum of two Gaussian fits (dark grey lines) in case of
    slit\,1b). The residuals are shown in dark grey. {\bf
    Middle left panel:} the pv diagram showing the \Ha\ velocities  extracted
    from the archival spectrum; {\bf lower panels:} the pv diagrams of the
    three spectra that were obtained by us. The solid lines represent the \HI\
    velocities extracted from the velocity map of the main component (see
    Fig.~\ref{n5408.hi.decomp.hres}, middle panel).}
\label{n5408pv}
\end{figure*}

We then performed a Gaussian decomposition of the \Ha\ line profiles as
measured from the long-slit spectra. We measured the peak velocities in order
to create position-velocity (pv) diagrams (Figs.~\ref{ic4662pv}, lower row and
\ref{n5408pv}, middle left panel and lower row). The \Ha\ velocities are
indicated by + symbols. The corresponding \HI\ velocities (solid lines)
were extracted from the peak velocity fields of the main component
(Figs.~\ref{ic4662.hi.decomp} and \ref{n5408.hi.decomp.hres}, middle
panels). Both the \HI\ and the \Ha\ velocities are not corrected for
inclination. Figures~\ref{ic4662pv} and \ref{n5408pv} also show the slit
positions, plotted over the \Ha\ image of each galaxy. Position\,0 in the pv
diagrams is marked by a small circle; the arrows indicate increasing distance
from 0 in a positive sense. Examples illustrating the quality of the spectra
and the analysis are given for both data sets in the upper right panel of
Fig.~\ref{ic4662pv} and in the upper and middle right panels of
Fig.~\ref{n5408pv}.

At first glance, the \Ha\ emission line is quite broad along all
slits. Observations of the ionised gas in other dwarf galaxies revealed
outflows with expansion velocities of about 20 to 50\skms\ \citep[e.g.,][]{Schwartz2004,vanEymeren2009a,vanEymeren2009c}. For the data used in our
analysis, this implies that it will be difficult to separate blended lines
because of the relatively low spectral resolution (60\skms\ for the archival
spectrum of NGC\,5408, 112\skms\ for all other spectra). Therefore, we fitted
the profiles with only one Gaussian unless a fit with two Gaussians gave
reasonable values, i.e., $FWHM_{\rm obs}>FWHM_{\rm instr}$. In many cases,
a single Gaussian fit led to large residuals, which indicates that the \Ha\
line is a superposition of at least two components (see, e.g.,
Fig.~\ref{ic4662pv}, upper right panel).

\begin{figure}
\centering
\includegraphics[width=.48\textwidth]{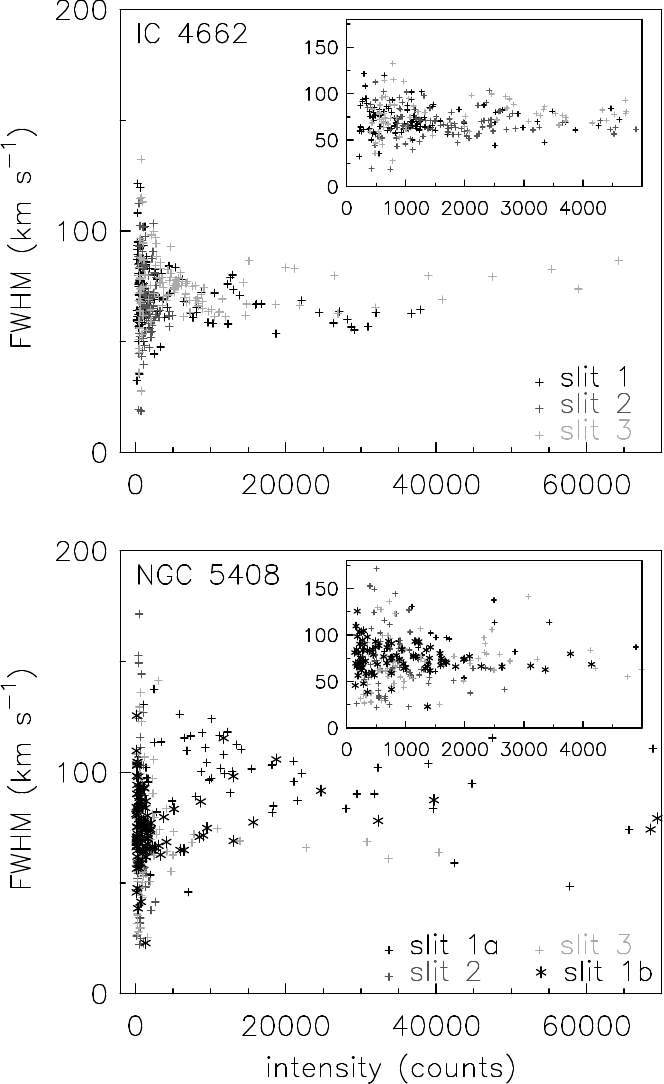}
\caption[]{The Gaussian \emph{FWHM} of the \Ha\ line, as measured from the
  long-slit spectra of both galaxies after correcting for instrumental
  broadening, \emph{vs.} the integrated intensity. The small panels in
  the upper right corner of both diagrams show an enlargement of the
  \emph{FWHMs} of the low intensity lines.}
\label{fwhm}
\end{figure}

In Fig.~\ref{fwhm} we present the \emph{FWHMs}, corrected for instrumental
broadening $FWHM^2_{corr}=FWHM^2_{obs}-FWHM^2_{instr}$, of all fitted lines
\emph{vs.} the integrated intensity. Different shades of grey indicate
\emph{FWHMs} measured in spectra from different slit positions. The black
stars denote the data of the archival spectrum of NGC\,5408.

In the following subsections, we describe the kinematics of the neutral and
ionised gas, point out peculiarities, and compare the kinematic behaviour of
both gas components.
\subsubsection{IC 4662}
\label{I4662longslit}
The high-resolution \HI\ velocity maps of IC\,4662 show that the neutral
  gas often consists of a single component that follows the overall
  rotation (Fig.~\ref{ic4662.hi.decomp}). Red-shifted gas can be seen in the filamentary north-eastern part of the optical disc of IC\,4662, which is
offset from the main component by about 55\skms. We did not find
any more outflowing neutral gas associated with the optical disc. However, we
detected blue-shifted gas north-west and south-east of the optical disc with
expansion velocities of about 40 to 50\skms\ with respect to the main
component, whose morphology gives the impression of some kind of bipolar
  outflow.

We then used the \HI\ velocity values of the main component as a reference
value for the velocities of the ionised gas. We obtained optical spectra of
IC\,4662 along three slit positions (Fig.~\ref{ic4662pv}, lower row). Slit~1
intersects the southern \HII\ region, the main \HII\ region complex, and the
diffuse gas in between. South of the main \HII\ region complex, i.e., from
$-160$\arcsec\ to $-80$\arcsec, the ionised gas roughly follows the \HI\
velocities. In most cases, however, the subtraction of the fitted Gaussian
from the observed profile reveals two residuals which are symmetrically
distributed. The \emph{FWHM} is measured to be 70\skms, which suggests a blue-
and a red-shifted component with a velocity offset of 35\skms\ with respect to
the \HI\ line. Between $-80$\arcsec\ and $-15$\arcsec, the ionised gas appears
to be slightly red-shifted by 20 to 30\skms\ with respect to the neutral gas.

Slit~2 intersects the northern part of the main \HII\ region complex. The \Ha\
emission detected along this slit follows the \HI\ velocities quite well apart
from a deviation of up to 20\skms\ red-shifted between 50\arcsec\ and
80\arcsec. The \emph{FWHM} of the lines in this area is quite low in
comparison to the other positions along the slit, and the residuals after
subtracting a single Gaussian are negligible. That means that here, we
directly detected expanding gas located in the north-east of the optical
disc. 

Slit~3 intersects the central part of the main \HII\ region complex. The \Ha\
velocities observed along this slit are in good agreement with the \HI\
velocities from 50\arcsec\ to 150\arcsec. We measured a \emph{FWHM} of about
70\skms\ and detected significant residuals after subtracting a single
Gaussian (see, e.g., Fig.~\ref{ic4662pv}, upper right panel). Therefore,
the ionised gas probably consists of two components, a blue- and a red-shifted
one, expanding with 35\skms\ with respect to the neutral gas. In the eastern filamentary area, i.e., from 0\arcsec\ to
50\arcsec, the pv-diagram shows a significant deviation of the \Ha\ velocities
from the \HI\ velocities. While the \Ha\ velocities still seem to follow the
velocity gradient of the neutral gas between 50\arcsec\ and 150\arcsec, the \HI\
velocities drop by about 40\skms. Looking at the velocity map of the main
component of the neutral gas (see Fig.~\ref{ic4662.hi.decomp}, middle panel),
it becomes clear that this drop is part of the large-scale velocity field and
coincides with the twist in the central parts where the velocity gradient
changes its direction (see also Sect.~\ref{SectI4662morpho}). That indicates
that the neutral and ionised gas are kinematically decoupled. Still, if we
extend the gradient between 50\arcsec\ and 150\arcsec\ down to 0\arcsec, the
ionised gas seems to be red-shifted by 20 to 30\skms. This red-shifted gas
component was also detected in \HI\ (see above).

Summarised, it can be said that the neutral and ionised gas are mostly in good
agreement. The \Ha\ residuals often imply the existence of two components,
symmetrically distributed around the \HI\ reference value. There are a few
obvious discrepancies, which might indicate outflowing ionised gas, but which
have also to be discussed in the context of a very disturbed \HI\ velocity
field (see Sect.~\ref{Sectouflows}).
\subsubsection{NGC 5408}
\label{N5408longslit}
Similar to IC\,4662, the \HI\ line profiles of NGC\,5408 often reveal a
  single component that follows the overall rotation. Coinciding with the
  northern part of the optical galaxy, the \HI\ shows an additional component,
  which is blue-shifted with respect to the main line by about 30\skms\ (see
  Fig.~\ref{n5408.hi.decomp.hres}).

Again, we used the \HI\ peak velocities of the main component as a reference
value for the \Ha\ velocities. We analysed the kinematics of the ionised gas within NGC\,5408 using spectra
provided by two different observing runs (see Sect.~\ref{medspec}). The panels
located in the lower row of Fig.~\ref{n5408pv} show the pv diagrams of our own
data, while the middle left panel represents the archival data.

In comparison to IC\,4662, the velocities of NGC\,5408 seem to deviate more
and stronger from the \HI\ reference value. Although the \Ha\ line profile is
generally well fitted by a single Gaussian, we detected more outflows as the
outflowing gas often seems to dominate the intensity distribution.

The centre of the \Ha\ emission in the south-west harbours a strong
blue-shifted component, which has a velocity offset of about 60\skms\ in
comparison to the neutral gas (see slit~1a). In slit~1b (the archival
data), there is a hint of even higher blue-shifted ionised gas (up
to 100\skms). However, as only three measurements of this fast expanding
outflow are available, we will not further consider it. Additionally,
slit~1b shows a red-shifted component with a similar offset as seen in
slit\,1a of about 50\skms. This component is only visible in the higher
resolution data, although a few measurements have been made in slit~1a as well
(at $-60$\arcsec).

The southern part of the shell structure in the east (see
Sect.~\ref{SectN5408morpho}) can kinematically be detected in slit~1b, 2, and
3. Slit~1b, that covers the edge of the shell, shows a blue-shifted component
with a velocity offset of about 20\skms\ (at $-80$\arcsec) with respect to the
\HI\ reference value. In slit~2, which fully intersects the potential
superbubble, the blue-shifted gas is offset by 30\skms\ at
30\arcsec. Additionally, we see red-shifted gas at 20\skms\ along the whole
slit and a hint of an even higher red-shifted outflow of 70\skms\ at
40\arcsec. Slit~3, that also fully intersects the potential superbubble, but
at a different angle, shows blue-shifted gas at 30\arcsec, offset by
50\skms, and at 80\arcsec, offset by 30\skms. The latter outflow coincides
with the northern part of the shell. Red-shifted gas at 20\skms\ is visible
between the two blue-shifted outflows.

As mentioned above, the neutral gas shows a hint of a blue-shifted
  component in the northern part of the optical disc, offset by about 30\skms\
  from the main component. This might be the counterpart of the expanding
  ionised gas. Other than that, the neutral gas does not seem to follow the
  motions of the ionised gas.

Altogether, the kinematic analysis confirms the suggestion we made in
Sect.~\ref{SectN5408morpho}, i.e., that we detected an expanding superbubble
in the north-east of NGC\,5408. However, the relatively low spectral
resolution and the poor spatial coverage make it impossible to model this
superbubble.
\section{Discussion}
Our analysis has shown that both galaxies show differences in the kinematic
behaviour of the neutral and ionised gas components. Often, this can be
interpreted as evidence for outflows of the ionised gas. Sometimes, these
outflows are detected in both \Ha\ and \HI. In this section, we discuss our
results and make some predictions about the fate of the gas. We will also
assess our results with respect to other studies of dwarf galaxies. Finally,
we will come back to the peculiar \HII\ region in the south of IC\,4662.
\subsection{The outflows and the special case of IC\,4662}
\label{Sectouflows}
The outflow velocities vary from 20 up to 60\skms\ in both galaxies. In
  some cases we found an \HI\ counterpart for an expanding ionised gas
  structure, moving at about the same velocity. NGC\,5408 shows little
  evidence for line-splitting in \HI. The neutral gas distribution seems to be
  warped in the outer parts, which might indicate gas accretion, interaction
  with gas-rich companions or infall of \HI\ clouds \citep[see ][and references
  therein]{Sancisi2008}. However, NGC\,5408 appears to be quite isolated and
  no \HI\ clouds could be found in its near vicinity on our deep \HI\
  maps. Furthermore, gas infall would result in a lopsided gas distribution,
  i.e., more gas on one side than on the other, which is definitely not seen
  in NGC\,5408. Therefore, the outflow scenario is much more likely.

  From the morphology and the kinematics, we suggest that NGC\,5408 harbours
  at least one expanding superbubble. In comparison to IC\,4662, this galaxy
  shows outflows almost everywhere across the disc.

For IC\,4662 no XMM Newton observations are available and the ROSAT X-ray
  All-Sky Survey does not show a source that could be associated with this
  galaxy. Nevertheless, IC\,4662 is a very complex system: the comparison of
  the neutral and ionised gas kinematics revealed that it is very difficult to take the \HI\ velocities as a reference value for
the \Ha\ velocities as the \HI\ velocity field is quite distorted. Especially
the discrepancy in slit\,3 (see Fig.~\ref{ic4662pv}, lower right panel) is
  probably due to the large-scale \HI\ velocity field and not due to an outflow of the ionised gas. The same might be true for
slit\,2 (between 30\arcsec\ and 70\arcsec) where the \HI\ velocities again
drop significantly. However, here, the ionised gas seems to follow that
movement. If we just look at the velocities of the ionised gas and try to
define a velocity gradient across each slit, a velocity offset of about
20\skms\ blue-shifted can be seen, which coincides with the north-eastern
filamentary area. From the morphology alone expanding gas is expected, but
because of the distorted \HI\ velocity field, and the lack of spatial coverage
  and spectral resolution of the long-slit spectra, it is difficult to
  quantify the outflow velocities.

As can be seen very nicely in NGC\,5408 (Fig.~\ref{n5408pv}, lower row), the
kinematics of the ionised gas are dominated by stellar feedback causing
offsets from the rotation velocity of the galaxy. This implies that the \Ha\
velocities can not be assumed to reflect the rotation of a galaxy; instead, we
rely on the \HI\ velocities. However, in the case of IC\,4662, the \HI\
velocity field is very distorted as well, which is why we are very careful
about our detections.

Looking at the large-scale structure, IC\,4662 shows line-splitting of the
\HI\ line at many positions across the galaxy (Fig.~\ref{IC4662profiles}). As
already mentioned above (Sect.~\ref{I4662HIkinematics}), the two components
are often separated by about 70\skms. Given the size of the optical galaxy in
comparison to the distribution of the neutral gas, this is probably not caused
by an interaction of stars and the ISM. As we see global line-splitting,
  the infall of single \HI\ clouds is also highly unlikely. Instead, the
line-splitting is probably a consequence of a large scale distortion like,
e.g., a merger process. As IC\,4662 is a very isolated galaxy, we may rule out
any kind of current tidal interaction with another object. We think that it is
more likely that two dwarf galaxies collided in the past and are now in the
process of merging. This hypothesis explains the distorted kinematics we
observe in IC\,4662, the two perpendicular systems as observed in the \HI\
distribution, and also the \HI\ tail we found in the eastern part of the
galaxy (see Fig.~\ref{IC4662mom}). There is increasing evidence that
interactions and mergers of dwarf galaxies trigger star-formation activity in
dwarf objects, but such features can only be detected when optical and radio
observations are carefully compared \citep[e.g.,][]{Lopez-Sanchez2009}. 
\subsection{Large \emph{FWHMs} in IC\,4662 and NGC\,5408}
The \Ha\ line profiles of both galaxies show very large \emph{FWHMs} (see
Fig.~\ref{fwhm}). In the case of IC\,4662, the \emph{FWHMs} of the low
intensity profiles are at about 70\skms. Even for high intensity profiles,
this number does not decrease, which means that the high \emph{FWHMs} are not an effect of low S/N. 70\skms\
are far too high to be explained by thermal broadening alone as this would
require an electron temperature of 125000\degr! Taking a reasonable value of
the electron temperature of 10000\,K gives a contribution of the thermal component of about 23\skms. This still
leaves us with 66\skms\ that have to be explained by other mechanisms. As
shown in Sect.~\ref{I4662longslit}, the residuals after subtracting a single
Gaussian from the observed line profile usually indicate that the \Ha\ line is
a superposition of at least two components. As the line has a symmetric shape,
this may indicate expanding bubbles.

NGC\,5408 also shows quite high \emph{FWHMs}. Here, it becomes obvious that this
  result is also not an effect of the spectral resolution as the line widths
  remain equally broad when measured at higher spectral resolution (from the
  archival spectrum). In comparison to IC\,4662, the \emph{FWHMs} are more scattered, although they are preferably located in a
band centred at 70\skms, which is similar to the average value found for
IC\,4662. Additionally, the ionised gas observed in slit~1b reaches
\emph{FWHMs} of up to 100\skms. These high values have mainly been found
in the western \HII\ region, which is where the X-ray source is located.

Detailed studies of the velocity dispersion in \HII\ regions have been
  carried out by \citet{Munoz-Tunon1996} and later by
  \citet{Martinez-Delgado2007}. Their diagrams show very similar patterns
  compared to ours: a narrow horizontal band and inclined bands at higher
  velocity dispersion, which they interprete as coming from faint expanding
  bubbles and shells. Indeed, we found the largest \emph{FWHMs} to coincide
  with areas of filamentary structure (IC\,4662) or potential superbubbles
  (NGC\,5408).
\subsection{Outflow or galactic wind?}
As we could not derive a reasonable rotation curve from the distorted velocity
field of IC\,4662, we focus this analysis on NGC\,5408.

\begin{figure}
\centering
\includegraphics[width=.48\textwidth]{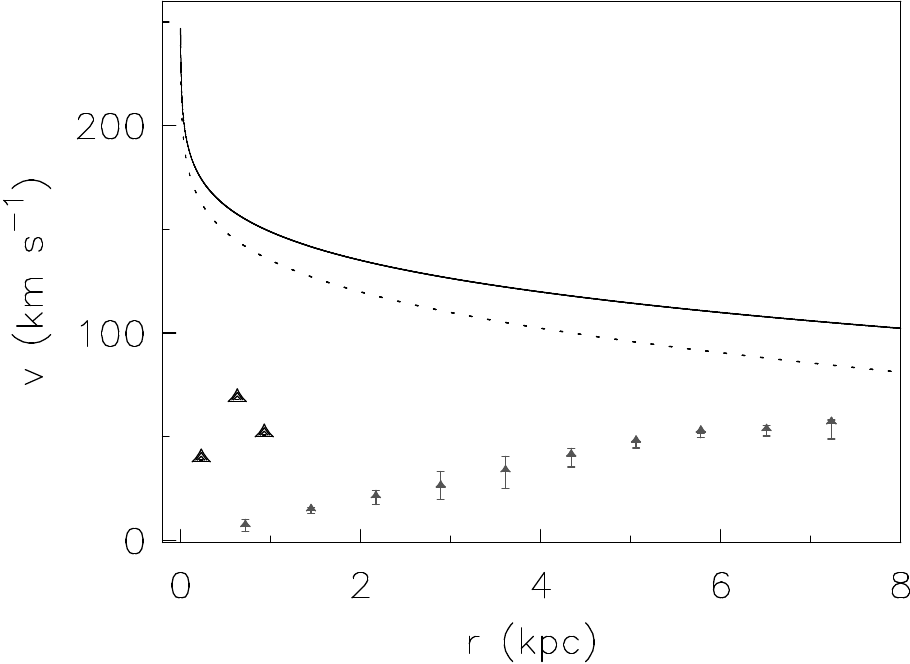}
\caption[Outflow or galactic wind?]{Outflow or galactic wind? The expansion
velocities of the outflowing gas structures in NGC\,5408 are compared with the
escape velocities calculated by using the pseudo-isothermal halo model. The
escape velocity is calculated for an isothermal halo of $r_{\rm
  max}\rm=9.5\,kpc$ (dotted line), which fits the size of the \HI\
distribution, and $r_{\rm max}\rm =19\,kpc$ (solid line). The observed
rotation curve is indicated by small grey triangles. The error bars represent
receding and approaching side. The expanding gas structures are marked by
black large triangles.}
\label{ICNGCesc}
\end{figure}

Following the method described in \citet{vanEymeren2009a}, we compared the
expansion velocities with the escape velocity of NGC\,5408 in order to make
predictions about the fate of the outflowing gas structures. The escape
velocity was estimated by using the cored pseudo-isothermal halo model,
which represents the shape of the rotation curves best, especially in the
inner few kpcs \citep[e.g.,][]{vanEymeren2009b}, which is where we detected
the outflows. Only the circular velocity and the maximum halo
radius $r_{\rm max}$ have to be known in order to calculate the escape
velocity. The circular velocity was estimated from the plateau of the \HI\
rotation curve. We derived the escape velocity for two different radii. First,
we took the outer radius of the \HI\ intensity distribution (9.5\,kpc) as a
lower limit for the halo size. As a second estimate we calculated the escape
velocity using twice the \HI\ radius (19\,kpc), which we consider to be a more
physical assumption. The resulting curves are shown in Fig.~\ref{ICNGCesc}
(dotted and solid lines respectively). The observed rotation curve is
indicated by small grey triangles; the outflows with their distances
from the dynamic centre and their velocities compared to the \HI\ velocities
are marked by large black triangles. A higher $r_{\rm max}$
results in an increase of the escape velocity. As the radius of the dark
matter halo $r_{\rm max}$ is probably much higher than assumed, the escape
velocities will be even higher than the ones we derived. The curve with
$r_{\rm max}=r_{\rm HI,max}$ gives therefore a lower limit for the escape
velocity values. As can be seen, the expansion velocities of the outflows,
although partially quite significant, stay far below the escape velocity, even
for the lower limit case. This is mainly due to the fact that they were found
close to the dynamic centre where the escape velocities are high.
\subsection{Comparison with previous work}
We now compare our results for IC\,4662 and NGC\,5408 with similar
  studies from the literature. Many dwarf galaxies reveal filaments and
  shell-like structures up to kpc-size on deep \Ha\ images \citep[see, e.g.,
  ][]{Hunter1993,Bomans1997,vanEymeren2007}. Spectral line data show \Ha\ line splitting,
  indicating outflowing gas, which is often associated with these filaments
  and shells \citep[e.g.,][]{Martin1998}. Partly, the neutral gas is visibly affected by the expanding
  ionised gas as can, e.g., be seen in the starburst dwarf galaxy NGC\,5253
  \citep[][]{Kobulnicky2008}. The neutral gas is pushed by the ionised gas,
  which leads to the observed expansion velocities. On the other hand, the
  surrounding neutral gas decelerates the expanding bubble so that it
  might eventually be stalled by the neutral gas envelope
  \citep[][]{Kobulnicky2008}. Figure~\ref{momhres} shows very nicely how the
  expanding gas can affect the neutral gas, as, e.g., NGC\,5408 has a local
  \HI\ minimum in the centre of the potential superbubble. A very similar
  morphology has been found in NGC\,5253 \citep{Lopez-Sanchez2010}. NGC\,2366
  \citep{vanEymeren2009a} and NGC\,4861 \citep{vanEymeren2009c} also show \HI\
  minima coinciding with expanding ionised gas structures.

The expansion velocities of the galaxies in our sample (including NGC\,2366
and NGC\,4861) vary between 20\skms\ and 60\skms. These values are in good
agreement with previous studies: \citet{Heckman1995} and
\citet{Westmoquette2008} found expansion velocities between 50 and 100\skms\
in NGC\,1569; \citet{Schwartz2004}, using absorption line studies, found the
expansion velocities of a sample of dwarf galaxies to be about 30\skms. An
exception seems to be the Wolf-Rayet blue compact dwarf galaxy He\,2-10, where
expansion velocities up to 250\skms\ have been detected
\citep[][]{Mendez1999}. Note that more massive galaxies like M\,82 show
outflows with expansion velocities higher than 100\skms\
\citep[][]{Schwartz2004,Strickland2007,Westmoquette2009}. In our sample, the
highest velocities have been found in NGC\,5408.

How high do the expansion velocities have to be to turn an outflow into a
galactic wind? Coming back to our sample (including NGC\,2366 and NGC\,4861,
but excluding IC\,4662 whose velocity field is too distorted to derive a
rotation curve and the associated escape velocity), it could be shown that the
expansion velocities of the outflows only account for 30 to 50\%\ of the
escape velocities. Note that
these velocities were calculated under the assumption that the radius of the
dark matter halo $r_{\rm max}$ equals the \HI\ radius. As $r_{\rm max}$
is probably much higher, the resulting escape velocities will be higher, which
therefore further decreases the possibility of a galactic wind. This result is
consistent with studies of other dwarf galaxies
\citep[e.g.,][]{Martin1998,Schwartz2004}. An exception is NGC\,1569 which rotates
very slowly with only 30\skms \citep{Heckman1995}, but harbours outflows with
expansion velocities of up to 100\skms\
\citep[][]{Heckman1995,Westmoquette2008}. Still, a galactic wind could not be
confirmed, although the chances are much higher than in other dwarf
galaxies. Another exception is He\,2-10 \citep{Mendez1999}, which however, as
mentioned before, shows very fast expanding gas with velocities up to
250\skms.

At first glance, all of these non-detections of a galactic wind look like a
contradiction with theoretical models like, e.g., the superbubble model by
\citet{MacLow1999}. Note, however, that outflowing gas is mostly detected close
to the dynamic centre, where the escape velocities are very high. Only a few
kpcs away from the dynamic centre, the escape velocity drops significantly
(see, e.g., Fig.~\ref{ICNGCesc}), increasing the chance of a galactic wind. As
the surface brightness goes with $n_e^2$ with $n_e$ being the electron
density, a less dense gas will be much more difficult to detect. This implies
that the outer regions of an outflow will easily escape the detection in
emission line studies \citep[e.g.,][]{Bomans2007}, which means that we
consistently miss the fastest moving gas. A study of the interstellar NaI
absorption line (which is as an absorption line not affected by the $n_e^2$
problem) does not help since NaI traces much cooler, denser gas, which is
probably not as much accelerated as the diffuse HII.

Additionally, we have to take into account the dynamic masses of our sample
galaxies, which are of the order of a few times 10$^9$\Msun, i.e., close to
the upper limit of the mass range of dwarf galaxies. As Fig.~2 in
\citet{MacLow1999} shows, at total masses of 10$^9$\Msun, most of the gas is
still bound to the disk when the energy input stops. This strong
  dependence of blow-out on the mass of the galaxy has already been noted by
\citet{DeYoung1994}. Furthermore, the central star clusters in, e.g.,
NGC\,2366 \citep{Drissen2001} or NGC\,4861 \citep[e.g.,][]{Barth1994} are only
a few Myrs old. If we assume that the current star formation episode is the
driver of the outflows, it makes a superbubble blowout even more unlikely.

It is also possible that the expansion velocity of the outflows will never get
close to or beyond the escape velocity due to the dark matter halo that slows
down the gas that was expelled from the disc \citep[][]{Silich1998}. Their 2d
calculations of multi-supernova remnants evolving in dwarf galaxies show that
the total retention of the outflowing gas is very likely for galaxies with an
ISM mass of about 10$^9$\Msun.

Altogether, this means that our observations (including NGC\,2366 and
NGC\,4861) are in good agreement with other observations and with the
simulations. All sample galaxies harbour filaments and shells that could be
shown to expand, but no evidence for a galactic wind could be found.

As the expansion velocities in IC\,4662 are comparable to those detected in
the other three galaxies, we suppose that we would find similar results for
this galaxy. However, if this galaxy is a merging system, processes with much
more impact than star formation activity probably dominate the kinematic
behaviour.
\subsection{The southern \HII\ region in IC\,4662}
\label{southHII}
The southern \HII\ region in IC\,4662 has been and still is a subject of great
interest. It appears to be detached from the main body. Furthermore,
\citet{Hidalgo-Games2001} pointed out that it differs significantly from the
main body in oxygen abundance. Therefore, the question arises whether
it is an own system interacting with IC\,4662 (e.g., H. Lee, private
communication) or just an unusually placed OB association. Our deep \Ha\ image
shows that this bright \HII\ region is at least connected to the main complex
by a chain of small \HII\ regions and diffuse filamentary gas structures
\citep[see Sect.~\ref{LVHISgenmorph} and ][]{Noeske2003}. Moreover, together
with the main body this region is embedded into the highest column density
\HI. Additionally, its kinematics do not differ from the main body, but follow
the rotation of the whole optical system. All these detections suggest that
the southern \HII\ region is part of IC\,4662.

Similar features have been found in the blue compact dwarf galaxy
  NGC\,1140: a chain of \HII\ regions that emanates south of the main body,
  and a core which is kinematically decoupled from the rest of the galaxy
\citep[][]{Hunter1994,Westmoquette2010}. Therefore, this galaxy was suggested
  to be a merger remnant \citep{Hunter1994}. A study of 20 starburst galaxies
  by \citet{Lopez-Sanchez2010b} has shown that in fact almost all of them show
  signs of interaction or of merging with small companions. They conclude that
  interactions play an important role in the triggering mechanism of the
  strong star-formation activity observed in dwarf galaxies.

Taking into account all the observational peculiarities and this similarity to NGC\,1140, we definitely need to discuss the detached \HII\ region in IC\,4662 in the
context of a possible merger. If IC\,4662 indeed consists of two merging or
merged systems, the difference in the chemical composition might indicate that
the main \HII\ region complex belongs to one galaxy and the southern \HII\
region to the second galaxy.  Detailed multi-wavelength observations of
IC\,4662 (and other blue compact dwarf galaxies) are on the way \citep[see
  also ][]{Lopez-Sanchez2009}.
\section{Summary}
We analysed \HI\ and optical data of the two peculiar dwarf galaxies IC\,4662
and NGC\,5408 to compare the morphology and the kinematics of the neutral and
ionised gas components.

In both galaxies, we detected outflows in \Ha, sometimes with an \HI\
counterpart, which show expansion velocities of 20 to 60\skms. The
\emph{FWHMs} of the \Ha\ lines are as high as 70\skms, independent of the
intensity of the lines. This suggests that other mechanisms apart from thermal
broadening are taking place. In many cases, the \Ha\ line is probably a
superposition of at least two components, which indicates that expanding
superbubbles might cause the high \emph{FWHMs}. Higher spectral resolution
data are necessary to study the outflows in more detail.

We compared the expansion velocities measured in NGC\,5408 with the escape
velocity of the galaxy and conclude that it is very unlikely that the gas will
be blown away. This is in agreement with many previous studies.

As IC\,4662 has a very distorted \HI\ velocity field, we did not perform
an analysis of the fate of the gas. Nevertheless, we detected outflowing
ionised gas. The neutral gas often consists of two
components, separated by 70\skms. We suggest that this galaxy is actually a
merger of two independent dwarf systems. The most peculiar feature in IC\,4662
is the southern \HII\ region, whose origin is still not clear. Although our
analysis indicates that it is probably a part of the galaxy, it may also be
the remnant of one of the two merged systems.
\section*{Acknowledgements}
The authors would like to thank the anonymous referee for his very constructive feedback which helped to significantly improve this paper.\\
This work was partly supported by the Deutsche Forschungsgesellschaft (DFG)
under the SFB 591, by the Research School of the Ruhr-Universit\"at Bochum, by
the Australia Telescope National Facility, CSIRO, and by the DAAD. We made
extensive use of NASA's Astrophysics Data System (ADS) Bibliographic Services
and the NASA/IPAC Extragalactic Database (NED) which is operated by the Jet
Propulsion Laboratory, California Institute of Technology, under contract with
the National Aeronautics and Space Administration.
\label{lastpage}
\bibliographystyle{mn2e}
\bibliography{bibliography}

\hyphenation{Post-Script Sprin-ger}
\begin{thebibliography}{}

\bibitem[\protect\citeauthoryear{Barth, Cepa, Vilchez \& Dottori}{Barth
  et~al.}{1994}]{Barth1994}
Barth C.~S.,  Cepa J.,  Vilchez J.~M.,    Dottori H.~A.,  1994, AJ, 108, 2069

\bibitem[\protect\citeauthoryear{{Begeman}}{{Begeman}}{1989}]{Begeman1989}
{Begeman} K.~G.,  1989, A\&A, 223, 47

\bibitem[\protect\citeauthoryear{Bohuski, Burbidge, Burbidge \& Smith}{Bohuski
  et~al.}{1972}]{Bohuski1972}
Bohuski T.~J.,  Burbidge E.~M.,  Burbidge G.~R.,    Smith M.~G.,  1972, ApJ,
  175, 329

\bibitem[\protect\citeauthoryear{Bomans}{Bomans}{2001}]{Bomans2001}
Bomans D.~J.,  2001, ApSS, 276, 783

\bibitem[\protect\citeauthoryear{Bomans}{Bomans}{2005}]{Bomans2005}
Bomans D.~J.,  2005, in AIP Conf. Proc. 783: The Evolution of Starbursts
  Outflows and galactic winds of dwarf galaxies.
p.~98

\bibitem[\protect\citeauthoryear{Bomans, Chu \& Hopp}{Bomans
  et~al.}{1997}]{Bomans1997}
Bomans D.~J.,  Chu Y.,    Hopp U.,  1997, AJ, 113, 1678

\bibitem[\protect\citeauthoryear{Bomans, van Eymeren, Dettmar, Weis \&
  Hopp}{Bomans et~al.}{2007}]{Bomans2007}
Bomans D.~J.,  van Eymeren J.,  Dettmar R.-J.,  Weis K.,    Hopp U.,  2007, New
  Astronomy Review, 51, 141

\bibitem[\protect\citeauthoryear{{de Blok}, {Walter}, {Brinks}, {Trachternach},
  {Oh} \& {Kennicutt}}{{de Blok} et~al.}{2008}]{deBlok2008}
{de Blok} W.~J.~G.,  {Walter} F.,  {Brinks} E.,  {Trachternach} C.,  {Oh}
  S.-H.,    {Kennicutt} R.~C.,  2008, AJ, 136, 2648

\bibitem[\protect\citeauthoryear{de
  Vaucouleurs}{de~Vaucouleurs}{1975}]{deVaucouleurs1975}
de Vaucouleurs G.,  1975, Nearby Groups of Galaxies.
Galaxies and the Universe, p.~557

\bibitem[\protect\citeauthoryear{de Vaucouleurs, de Vaucouleurs, Corwin, Buta,
  Paturel \& Fouque}{de~Vaucouleurs et~al.}{1991}]{deVaucouleurs1991}
de Vaucouleurs G.,  de Vaucouleurs A.,  Corwin H.~G.,  Buta R.~J.,  Paturel G.,
     Fouque P.,  1991, {Third Reference Catalogue of Bright Galaxies}.
Volume 1-3, XII, 2069 pp.~7 figs..~ Springer-Verlag Berlin Heidelberg New York

\bibitem[\protect\citeauthoryear{{De Young} \& {Heckman}}{{De Young} \&
  {Heckman}}{1994}]{DeYoung1994}
{De Young} D.~S.,  {Heckman} T.~M.,  1994, ApJ, 431, 598

\bibitem[\protect\citeauthoryear{Drissen, Crowther, Smith, Robert, Roy \&
  Hillier}{Drissen et~al.}{2001}]{Drissen2001}
Drissen L.,  Crowther P.~A.,  Smith L.~J.,  Robert C.,  Roy J.,    Hillier
  D.~J.,  2001, ApJ, 546, 484

\bibitem[\protect\citeauthoryear{Ferrara \& Tolstoy}{Ferrara \&
  Tolstoy}{2000}]{Ferrara2000}
Ferrara A.,  Tolstoy E.,  2000, MNRAS, 313, 291

\bibitem[\protect\citeauthoryear{{Heckman}, {Dahlem}, {Lehnert}, {Fabbiano},
  {Gilmore} \& {Waller}}{{Heckman} et~al.}{1995}]{Heckman1995}
{Heckman} T.~M.,  {Dahlem} M.,  {Lehnert} M.~D.,  {Fabbiano} G.,  {Gilmore} D.,
     {Waller} W.~H.,  1995, ApJ, 448, 98

\bibitem[\protect\citeauthoryear{Hidalgo-G{\'a}mez, Masegosa \&
  Olofsson}{Hidalgo-G{\'a}mez et~al.}{2001}]{Hidalgo-Games2001}
Hidalgo-G{\'a}mez A.~M.,  Masegosa J.,    Olofsson K.,  2001, A\&A, 376, 386

\bibitem[\protect\citeauthoryear{Hunter \& Gallagher}{Hunter \&
  Gallagher}{1997}]{Hunter1997}
Hunter D.~A.,  Gallagher J.~S.,  1997, ApJ, 475, 65

\bibitem[\protect\citeauthoryear{Hunter, Hawley \& Gallagher}{Hunter
  et~al.}{1993}]{Hunter1993}
Hunter D.~A.,  Hawley W.~N.,    Gallagher J.~S.,  1993, AJ, 106, 1797

\bibitem[\protect\citeauthoryear{{Hunter}, {van Woerden} \& {Gallagher}
  III}{{Hunter} et~al.}{1994}]{Hunter1994}
{Hunter} D.~A.,  {van Woerden} H.,    {Gallagher} III J.~S.,  1994, ApJS, 91,
  79

\bibitem[\protect\citeauthoryear{{Kaaret} \& {Corbel}}{{Kaaret} \&
  {Corbel}}{2009}]{Kaaret2009}
{Kaaret} P.,  {Corbel} S.,  2009, ApJ, 697, 950

\bibitem[\protect\citeauthoryear{Karachentsev, Dolphin, Tully, Sharina,
  Makarova, Makarov, Karachentseva, Sakai \& Shaya}{Karachentsev
  et~al.}{2006}]{Karachentsev2006}
Karachentsev I.~D.,  Dolphin A.,  Tully R.~B.,  Sharina M.,  Makarova L.,
  Makarov D.,  Karachentseva V.,  Sakai S.,    Shaya E.~J.,  2006, AJ, 131,
  1361

\bibitem[\protect\citeauthoryear{Karachentsev, Sharina, Dolphin, Grebel,
  Geisler, Guhathakurta, Hodge, Karachentseva, Sarajedini \&
  Seitzer}{Karachentsev et~al.}{2002}]{Karachentsev2002}
Karachentsev I.~D.,  Sharina M.~E.,  Dolphin A.~E.,  Grebel E.~K.,  Geisler D.,
   Guhathakurta P.,  Hodge P.~W.,  Karachentseva V.~E.,  Sarajedini A.,
  Seitzer P.,  2002, A\&A, 385, 21

\bibitem[\protect\citeauthoryear{{Kobulnicky} \& {Skillman}}{{Kobulnicky} \&
  {Skillman}}{2008}]{Kobulnicky2008}
{Kobulnicky} H.~A.,  {Skillman} E.~D.,  2008, AJ, 135, 527

\bibitem[\protect\citeauthoryear{Koribalski}{Koribalski}{2008}]{Koribalski2008}
Koribalski B.~S.,  2008, {The Local Volume HI Survey (LVHIS)}.
In Galaxies in the Local Volume, edited by Koribalski, B.~S.; Jerjen, H.,
  Berlin: Springer, 2008; p.~41

\bibitem[\protect\citeauthoryear{Koribalski et~al.,}{Koribalski
  et~al.}{2004}]{Koribalski2004}
Koribalski B.~S.,  et~al., 2004, AJ, 128, 16

\bibitem[\protect\citeauthoryear{{Koribalski} et~al.,}{{Koribalski}
  et~al.}{2010}]{Koribalski2010}
{Koribalski} B.~S.,  et~al., 2010, in prep.

\bibitem[\protect\citeauthoryear{{Lang}, {Kaaret}, {Corbel} \& {Mercer}}{{Lang}
  et~al.}{2007}]{Lang2007}
{Lang} C.~C.,  {Kaaret} P.,  {Corbel} S.,    {Mercer} A.,  2007, ApJ, 666, 79

\bibitem[\protect\citeauthoryear{{Larson}}{{Larson}}{1974}]{Larson1974}
{Larson} R.~B.,  1974, MNRAS, 169, 229

\bibitem[\protect\citeauthoryear{{Lauberts} \& {Valentijn}}{{Lauberts} \&
  {Valentijn}}{1989}]{Lauberts1989}
{Lauberts} A.,  {Valentijn} E.~A.,  1989, {The surface photometry catalogue of
  the ESO-Uppsala galaxies}.
Garching: European Southern Observatory, |c1989

\bibitem[\protect\citeauthoryear{{L{\'o}pez-S{\'a}nchez}}{{L{\'o}pez-S{\'a}nch%
ez}}{2010}]{Lopez-Sanchez2010b}
{L{\'o}pez-S{\'a}nchez} {\'A}.~R.,  2010, A\&A, in rev.

\bibitem[\protect\citeauthoryear{{L{\'o}pez-S{\'a}nchez} \&
  {Esteban}}{{L{\'o}pez-S{\'a}nchez} \& {Esteban}}{2008}]{Lopez-Sanchez2008a}
{L{\'o}pez-S{\'a}nchez} {\'A}.~R.,  {Esteban} C.,  2008, A\&A, 491, 131

\bibitem[\protect\citeauthoryear{{L{\'o}pez-S{\'a}nchez}, {Koribalski},
  {Esteban}, {Popping}, {van Eymeren} \& {Hibbard}}{{L{\'o}pez-S{\'a}nchez}
  et~al.}{2010}]{Lopez-Sanchez2009}
{L{\'o}pez-S{\'a}nchez} {\'A}.~R.,  {Koribalski} B.,  {Esteban} C.,  {Popping}
  A.,  {van Eymeren} J.,    {Hibbard} J.,  2010, {A multiwavelength analysis of
  Blue Compact Dwarf Galaxies: HI results}

\bibitem[\protect\citeauthoryear{{L{\'o}pez-S{\'a}nchez}, {Koribalski}, {van
  Eymeren} \& {Esteban}}{{L{\'o}pez-S{\'a}nchez}
  et~al.}{2010}]{Lopez-Sanchez2010}
{L{\'o}pez-S{\'a}nchez} {\'A}.~R.,  {Koribalski} B.,  {van Eymeren} J.,
  {Esteban} C. e.~a.,  2010, MNRAS

\bibitem[\protect\citeauthoryear{Mac~Low \& Ferrara}{Mac~Low \&
  Ferrara}{1999}]{MacLow1999}
Mac~Low M.,  Ferrara A.,  1999, ApJ, 513, 142

\bibitem[\protect\citeauthoryear{Marlowe, Heckman, Wyse \& Schommer}{Marlowe
  et~al.}{1995}]{Marlowe1995}
Marlowe A.~T.,  Heckman T.~M.,  Wyse R.~F.~G.,    Schommer R.,  1995, ApJ, 438,
  563

\bibitem[\protect\citeauthoryear{Martin}{Martin}{1998}]{Martin1998}
Martin C.~L.,  1998, ApJ, 506, 222

\bibitem[\protect\citeauthoryear{{Mart{\'{\i}}nez-Delgado}, {Tenorio-Tagle},
  {Mu{\~n}oz-Tu{\~n}{\'o}n}, {Moiseev} \&
  {Cair{\'o}s}}{{Mart{\'{\i}}nez-Delgado} et~al.}{2007}]{Martinez-Delgado2007}
{Mart{\'{\i}}nez-Delgado} I.,  {Tenorio-Tagle} G.,  {Mu{\~n}oz-Tu{\~n}{\'o}n}
  C.,  {Moiseev} A.~V.,    {Cair{\'o}s} L.~M.,  2007, AJ, 133, 2892

\bibitem[\protect\citeauthoryear{{M{\'e}ndez}, {Esteban}, {Filipovi{\'c} },
  {Ehle}, {Haberl}, {Pietsch} \& {Haynes}}{{M{\'e}ndez}
  et~al.}{1999}]{Mendez1999}
{M{\'e}ndez} D.~I.,  {Esteban} C.,  {Filipovi{\'c} } M.~D.,  {Ehle} M.,
  {Haberl} F.,  {Pietsch} W.,    {Haynes} R.~F.,  1999, A\&A, 349, 801

\bibitem[\protect\citeauthoryear{{Mu{\~n}oz-Tu{\~n}{\'o}n}, {Tenorio-Tagle},
  {Castaneda} \& {Terlevich}}{{Mu{\~n}oz-Tu{\~n}{\'o}n}
  et~al.}{1996}]{Munoz-Tunon1996}
{Mu{\~n}oz-Tu{\~n}{\'o}n} C.,  {Tenorio-Tagle} G.,  {Castaneda} H.~O.,
  {Terlevich} R.,  1996, AJ, 112, 1636

\bibitem[\protect\citeauthoryear{{Noeske}, {Papaderos}, {Cair{\'o}s} \&
  {Fricke}}{{Noeske} et~al.}{2003}]{Noeske2003}
{Noeske} K.~G.,  {Papaderos} P.,  {Cair{\'o}s} L.~M.,    {Fricke} K.~J.,  2003,
  A\&A, 410, 481

\bibitem[\protect\citeauthoryear{Norman \& Ikeuchi}{Norman \&
  Ikeuchi}{1989}]{Norman1989}
Norman C.~A.,  Ikeuchi S.,  1989, ApJ, 345, 372

\bibitem[\protect\citeauthoryear{Richter, Lorenz, Bohm \& Priebe}{Richter
  et~al.}{1991}]{Richter1991}
Richter G.~M.,  Lorenz H.,  Bohm P.,    Priebe A.,  1991, Astronomische
  Nachrichten, 312, 345

\bibitem[\protect\citeauthoryear{{Sancisi}, {Fraternali}, {Oosterloo} \& {van
  der Hulst}}{{Sancisi} et~al.}{2008}]{Sancisi2008}
{Sancisi} R.,  {Fraternali} F.,  {Oosterloo} T.,    {van der Hulst} T.,  2008,
  A\&ARv, 15, 189

\bibitem[\protect\citeauthoryear{{Sault}, {Teuben} \& {Wright}}{{Sault}
  et~al.}{1995}]{Sault1995}
{Sault} R.~J.,  {Teuben} P.~J.,    {Wright} M.~C.~H.,  1995, in {R.~A.~Shaw,
  H.~E.~Payne, \& J.~J.~E.~Hayes} ed., Astronomical Data Analysis Software and
  Systems IV Vol.~77 of Astronomical Society of the Pacific Conference Series,
  {A Retrospective View of MIRIAD}.
pp 433

\bibitem[\protect\citeauthoryear{{Schlegel}, {Finkbeiner} \&
  {Davis}}{{Schlegel} et~al.}{1998}]{Schlegel1998}
{Schlegel} D.~J.,  {Finkbeiner} D.~P.,    {Davis} M.,  1998, ApJ, 500, 525

\bibitem[\protect\citeauthoryear{{Schwartz} \& {Martin}}{{Schwartz} \&
  {Martin}}{2004}]{Schwartz2004}
{Schwartz} C.~M.,  {Martin} C.~L.,  2004, ApJ, 610, 201

\bibitem[\protect\citeauthoryear{Shapiro \& Field}{Shapiro \&
  Field}{1976}]{Shapiro1976}
Shapiro P.~R.,  Field G.~B.,  1976, ApJ, 205, 762

\bibitem[\protect\citeauthoryear{Silich \& Tenorio-Tagle}{Silich \&
  Tenorio-Tagle}{2001}]{Silich2001}
Silich S.,  Tenorio-Tagle G.,  2001, ApJ, 552, 91

\bibitem[\protect\citeauthoryear{Silich \& Tenorio-Tagle}{Silich \&
  Tenorio-Tagle}{1998}]{Silich1998}
Silich S.~A.,  Tenorio-Tagle G.,  1998, MNRAS, 299, 249

\bibitem[\protect\citeauthoryear{{Soria}, {Fender}, {Hannikainen}, {Read} \&
  {Stevens}}{{Soria} et~al.}{2006}]{Soria2006}
{Soria} R.,  {Fender} R.~P.,  {Hannikainen} D.~C.,  {Read} A.~M.,    {Stevens}
  I.~R.,  2006, MNRAS, 368, 1527

\bibitem[\protect\citeauthoryear{{Soria}, {Motch}, {Read} \& {Stevens}}{{Soria}
  et~al.}{2004}]{Soria2004}
{Soria} R.,  {Motch} C.,  {Read} A.~M.,    {Stevens} I.~R.,  2004, A\&A, 423,
  955

\bibitem[\protect\citeauthoryear{{Strickland} \& {Heckman}}{{Strickland} \&
  {Heckman}}{2007}]{Strickland2007}
{Strickland} D.~K.,  {Heckman} T.~M.,  2007, ApJ, 658, 258

\bibitem[\protect\citeauthoryear{{Strohmayer} \& {Mushotzky}}{{Strohmayer} \&
  {Mushotzky}}{2009}]{Strohmayer2009}
{Strohmayer} T.~E.,  {Mushotzky} R.~F.,  2009, ApJ, 703, 1386

\bibitem[\protect\citeauthoryear{{Strohmayer}, {Mushotzky}, {Winter}, {Soria},
  {Uttley} \& {Cropper}}{{Strohmayer} et~al.}{2007}]{Strohmayer2007}
{Strohmayer} T.~E.,  {Mushotzky} R.~F.,  {Winter} L.,  {Soria} R.,  {Uttley}
  P.,    {Cropper} M.,  2007, ApJ, 660, 580

\bibitem[\protect\citeauthoryear{{Tody}}{{Tody}}{1993}]{Tody1993}
{Tody} D.,  1993, in {R.~J.~Hanisch, R.~J.~V.~Brissenden, \& J.~Barnes} ed.,
  Astronomical Data Analysis Software and Systems II Vol.~52 of Astronomical
  Society of the Pacific Conference Series, {IRAF in the Nineties}.
pp 173

\bibitem[\protect\citeauthoryear{{van der Hulst}, {Terlouw}, {Begeman},
  {Zwitser} \& {Roelfsema}}{{van der Hulst} et~al.}{1992}]{vanderHulst1992}
{van der Hulst} J.~M.,  {Terlouw} J.~P.,  {Begeman} K.~G.,  {Zwitser} W.,
  {Roelfsema} P.~R.,  1992, in {Worrall} D.~M.,  {Biemesderfer} C.,   {Barnes}
  J.,  eds, Astronomical Data Analysis Software and Systems I Vol.~25 of
  Astronomical Society of the Pacific Conference Series, {The Groningen Image
  Processing SYstem, GIPSY}.
pp 131

\bibitem[\protect\citeauthoryear{{van Dokkum}}{{van
  Dokkum}}{2001}]{vanDokkum2001}
{van Dokkum} P.~G.,  2001, PASP, 113, 1420

\bibitem[\protect\citeauthoryear{van Eymeren}{van
  Eymeren}{2008}]{vanEymeren2008PhD}
van Eymeren J.,  2008, PhD thesis, Astronomisches Institut der
  Ruhr-Universitaet Bochum, Germany

\bibitem[\protect\citeauthoryear{van Eymeren, Bomans, Weis \& Dettmar}{van
  Eymeren et~al.}{2007}]{vanEymeren2007}
van Eymeren J.,  Bomans D.~J.,  Weis K.,    Dettmar R.-J.,  2007, A\&A, 474, 67

\bibitem[\protect\citeauthoryear{{van Eymeren}, {Marcelin}, {Koribalski},
  {Dettmar}, {Bomans}, {Gach} \& {Balard}}{{van Eymeren}
  et~al.}{2009a}]{vanEymeren2009a}
{van Eymeren} J.,  {Marcelin} M.,  {Koribalski} B.,  {Dettmar} R.-J.,  {Bomans}
  D.~J.,  {Gach} J.-L.,    {Balard} P.,  2009a, A\&A, 493, 511

\bibitem[\protect\citeauthoryear{{van Eymeren}, {Marcelin}, {Koribalski},
  {Dettmar}, {Bomans}, {Gach} \& {Balard}}{{van Eymeren}
  et~al.}{2009b}]{vanEymeren2009c}
{van Eymeren} J.,  {Marcelin} M.,  {Koribalski} B.~S.,  {Dettmar} R.,  {Bomans}
  D.~J.,  {Gach} J.,    {Balard} P.,  2009b, A\&A, 505, 105

\bibitem[\protect\citeauthoryear{{van Eymeren}, {Trachternach}, {Koribalski} \&
  {Dettmar}}{{van Eymeren} et~al.}{2009}]{vanEymeren2009b}
{van Eymeren} J.,  {Trachternach} C.,  {Koribalski} B.~S.,    {Dettmar} R.,
  2009, A\&A, 505, 1

\bibitem[\protect\citeauthoryear{{Westmoquette}, {Gallagher} \& {de
  Poitiers}}{{Westmoquette} et~al.}{2010}]{Westmoquette2010}
{Westmoquette} M.~S.,  {Gallagher} J.~S.,    {de Poitiers} L.,  2010, MNRAS, pp
  173

\bibitem[\protect\citeauthoryear{{Westmoquette}, {Smith} \&
  {Gallagher}}{{Westmoquette} et~al.}{2008}]{Westmoquette2008}
{Westmoquette} M.~S.,  {Smith} L.~J.,    {Gallagher} J.~S.,  2008, MNRAS, 383,
  864

\bibitem[\protect\citeauthoryear{{Westmoquette}, {Smith}, {Gallagher},
  {Trancho}, {Bastian} \& {Konstantopoulos}}{{Westmoquette}
  et~al.}{2009}]{Westmoquette2009}
{Westmoquette} M.~S.,  {Smith} L.~J.,  {Gallagher} J.~S.,  {Trancho} G.,
  {Bastian} N.,    {Konstantopoulos} I.~S.,  2009, ApJ, 696, 192

\bibitem[\protect\citeauthoryear{Yang, Chu, Skillman \& Terlevich}{Yang
  et~al.}{1996}]{Yang1996}
Yang H.,  Chu Y.-H.,  Skillman E.~D.,    Terlevich R.,  1996, AJ, 112, 146

\end{thebibliography}
\appendix
\section{Intensity-weighted mean \emph{vs.} Hermite velocity field}
\begin{figure*}
\centering
\includegraphics[width=\textwidth]{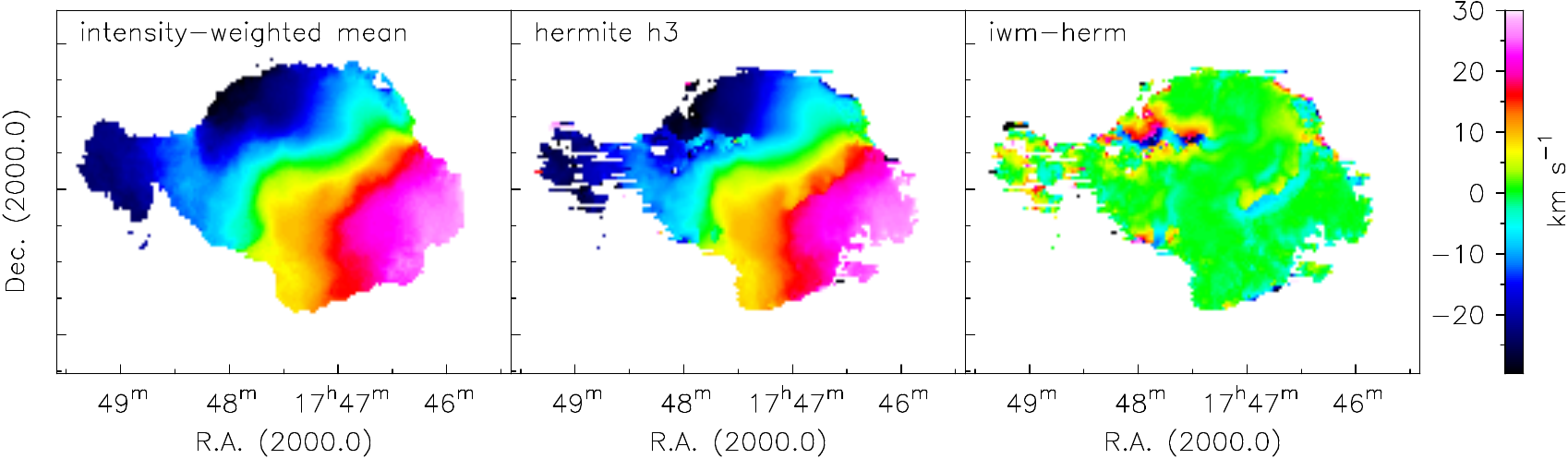}
\caption{A comparison of the intensity-weighted mean (iwm) velocity field (left
  panel) \emph{vs.} the Hermite h3 velocity field (middle panel). The
  residuals after subtracting the Hermite from the iwm velocity field is shown
  on the right panel. 
}
\label{IC4662iwm-herm}
\end{figure*}

Here, we present the intensity-weighted mean (iwm) and the Hermite velocity
field of IC\,4662 (Fig.~\ref{IC4662iwm-herm}, left and middle panels). By
applying Gauss-Hermite h3 polynomials to a set of spectra, the peak velocities
can be defined more accurately as this method is not biased by low S/N or
blended lines. However, in the case of IC\,4662 (and also in the case of
NGC\,5408), the \HI\ profiles are either symmetric and simple or the
components can well be differentiated. This is confirmed if we subtract the
Hermite velocity field from the iwm velocity field (Fig.~\ref{IC4662iwm-herm},
right panel): the two velocity fields only differ in areas of very low
S/N. The large scale velocity field remains the same. 
\end{document}